\documentclass[lettersize,journal]{IEEEtran}
\usepackage[utf8]{inputenc}
\usepackage[T1]{fontenc}
\usepackage{amsmath,amsfonts}
\usepackage{algorithmic}
\usepackage{algorithm}
\usepackage{array}
\usepackage[caption=false,font=normalsize,labelfont=sf,textfont=sf]{subfig}
\usepackage{textcomp} 
\newcommand{\textapprox}{\raisebox{0.5ex}{\texttildelow}} 
\usepackage{stfloats}
\usepackage{url}
\usepackage{verbatim}
\usepackage{graphicx}
\usepackage[format=plain,font=footnotesize,justification=justified,singlelinecheck=false,skip=0pt]{caption}
\usepackage[style=ieee,sorting=none,doi=false,isbn=false,url=false,eprint=false]{biblatex}
\usepackage{color,soul} 
\AtEveryBibitem{\clearfield{note}} 
\makeatletter 
\newcommand\semiHuge{\@setfontsize\semiHuge{22.72}{27.38}}
\makeatother

\begin{document}

\title{\semiHuge A Wireless, Multicolor Fluorescence Image Sensor Implant for Real-Time Monitoring in Cancer Therapy}

\author{Micah Roschelle*,~\IEEEmembership{Member,~IEEE}, Rozhan Rabbani*,~\IEEEmembership{Member,~IEEE,} Surin Gweon,~\IEEEmembership{Member,~IEEE,} Rohan Kumar, ~\IEEEmembership{Member,~IEEE,} Alec Vercruysse,~\IEEEmembership{Member,~IEEE,} Nam Woo Cho, Matthew H. Spitzer, Ali M. Niknejad,~\IEEEmembership{Fellow,~IEEE,} Vladimir M. Stojanovi\'{c},~\IEEEmembership{Fellow,~IEEE,} Mekhail Anwar,~\IEEEmembership{Member,~IEEE}
\thanks{This work was supported by the Office of the Director and the National Institute of Dental and Craniofacial Research of the National Institutes of Health under Award DP2DE030713 and the John V. Carbone Jr. Pancreatic Cancer Research Memorial Fund. \textit{(Corresponding authors: Micah Roschelle and Mekhail Anwar.)} *Equally contributing authors.

Micah Roschelle, Rozhan Rabbani, Surin Gweon, Rohan Kumar, Alec Vercruysse, Ali Niknejad, and Vladimir Stojanovi\'{c} are with the Department of Electrical Engineering and Computer Sciences, University of California at Berkeley, Berkeley CA 94720 USA. (email: micah.roschelle@berkeley.edu)

Nam Woo Cho is with the Department of Radiation Oncology and the Department of Otolaryngology-Head and Neck Surgery, University of California, San Francisco, CA 94158 USA.

Matthew Spitzer is with the Department of Otolaryngology-Head and Neck Surgery and the Department of Microbiology and Immunology, University of California, San Francisco, CA 94158 USA.

Mekhail Anwar is with the Department of Electrical Engineering and Computer Sciences, University of California at Berkeley, Berkeley, CA 94720 USA and also the Department of Radiation Oncology, University of California, San Francisco, CA 94158 USA. (email: mekhail@berkeley.edu, mekhail.anwar@ucsf.edu).
}
}



\maketitle
\begin{abstract}
Real-time monitoring of dynamic biological processes in the body is critical to understanding disease progression and treatment response. This data, for instance, can help address the lower than 50\% response rates to cancer immunotherapy. However, current clinical imaging modalities lack the molecular contrast, resolution, and chronic usability for rapid and accurate response assessments. Here, we present a fully wireless image sensor featuring a 2.5×5 mm\textsuperscript{2} CMOS integrated circuit for multicolor fluorescence imaging deep in tissue. The sensor operates wirelessly via ultrasound (US) at 5 cm depth in oil, harvesting energy with 221 mW/cm\textsuperscript{2} incident US power density (31\% of FDA limits) and backscattering data at 13 kbps with a bit error rate <10\textsuperscript{-6}. In-situ fluorescence excitation is provided by micro-laser diodes controlled with a programmable on-chip driver. An optical frontend combining a multi-bandpass interference filter and a fiber optic plate provides >6 OD excitation blocking and enables three-color imaging for detecting multiple cell types. A 36×40-pixel array captures images with <125 {\textmu}m resolution. We demonstrate wireless, dual-color fluorescence imaging of both effector and suppressor immune cells in \textit{ex vivo} mouse tumor samples with and without immunotherapy. These results show promise for providing rapid insight into therapeutic response and resistance, guiding personalized medicine.
\end{abstract}


\section{Introduction}
\IEEEPARstart{W}{IRELESS} 
, miniaturized, implantable sensors can monitor intricate biological processes unfolding in the body in real-time. Typically accessible only through highly invasive techniques, this data is crucial for advancing personalized medicine, tailoring treatments to individual responses to address the wide heterogeneity in therapeutic outcomes among patients.

One meaningful application is monitoring tumor response to cancer immunotherapy, a promising treatment that unlocks the patient's own immune system to fight cancer. For instance, immune checkpoint inhibitors (ICIs), a class of immunotherapy, have been shown to nearly double patient survival rates in melanoma \cite{hodi_f_stephen_improved_2010} and metastatic lung cancer \cite{reck_martin_pembrolizumab_2016} with a lower incidence of adverse effects compared to conventional treatments like chemotherapy \cite{gadgeel_updated_2020}. While more than 40\% of US cancer patients are estimated to be eligible for ICIs \cite{haslam_estimation_2019}, these therapies face a significant challenge: across most cancer types, less than 30\% of patients respond to treatment \cite{morad_hallmarks_2021, das_immune-related_2019}. For non-responders, time spent on ineffective therapies not only allows for their cancer to grow and spread, but also exposes them to unnecessary toxicity with high-grade adverse events rates often exceeding 10\% \cite{morad_hallmarks_2021} and financial burdens of more than \$150,000 per year \cite{verma_systematic_2018, chiang_cost-effectiveness_2021}. Rapid assessments of therapeutic response that also provide insight into the underlying mechanisms of resistance can help clinicians quickly identify non-responders and pivot to more effective second-line therapies to overcome resistance. However, such an assessment must capture the complex and dynamic interplay between various effector and suppressor immune cells and cancer that determines response \cite{morad_hallmarks_2021}. 

Current clinical imaging falls short of this goal. Anatomical imaging modalities such as computed tomography (CT) and magnetic resonance imaging (MRI) capture changes in tumor size, which take months to manifest and do not reliably correlate with response \cite{chai_challenges_2020}. These limitations are apparent in standard response criteria. For example, iRECIST defines a partial response as at least a 30\% reduction in tumor dimensions with a minimum size of 1 cm and recommends confirmation of disease progression at long 4--8 week intervals \cite{nishino_imaging_2019}, \cite{seymour_irecist_2017}. Alternatively, positron emission tomography (PET) can image the underlying biology with molecular contrast \cite{unterrainer_petct_2020}, but is fundamentally limited to imaging a single cell type or biomarker \cite{pratt_simultaneous_2023} at millimeter-scale resolution \cite{moses_fundamental_2011}. As the immune response depends on interactions between a variety of immune cells, it cannot be reliably predicted by a single biomarker \cite{gibney_predictive_2016, yang_liquid_2023}. Moreover, this millimeter-scale resolution averages out the spatial distributions of different cell populations within the tumor, shown to be increasingly important in understanding therapeutic resistance \cite{gohil_applying_2021, vitale_intratumoral_2021}.

Fluorescence microscopy, on the other hand, provides multi-cellular resolution across multiple biomarkers, essential to visualizing a more complete picture of the immune response. In fluorescence microscopy, targeted cells are labeled with fluorescent dyes, or fluorophores, which absorb light near a specific wavelength and emit light at slightly longer wavelengths \cite{lichtman_fluorescence_2005}. Multiple cell types can be imaged simultaneously by labeling each with a different color fluorophore. However, \textit{in vivo} optical imaging is constrained by scattering in tissue which fundamentally limits the penetration depth of light in the body to a few millimeters, even at near-infrared (NIR) wavelengths where tissue absorption is minimal and scattering is reduced \cite{owens_nir_2015}. Therefore, chronic fluorescence imaging at depth requires implantable imagers with integrated light sources providing in-situ illumination. 

Fluorescence imagers can be miniaturized to the scale of a single chip by eliminating bulky lenses through contact imaging \cite{papageorgiou_chip-scale_2020,aghlmand_65-nm_2023, zhu_ingestible_2023, moazeni_mechanically_2021, rustami_needle-type_2020}. To this end, prior work has demonstrated on-chip or in-package integration of focusing optics \cite{papageorgiou_chip-scale_2020, choi_fully_2020} as well as fluorescence filters \cite{aghlmand_65-nm_2023, zhu_ingestible_2023, moazeni_mechanically_2021, rustami_needle-type_2020, papageorgiou_angle-insensitive_2018} and light sources \cite{moazeni_mechanically_2021}. However, these systems are wired, precluding long-term implantation without risk of infection. While a fluorescence sensor with wireless radio-frequency (RF) communication is presented in \cite{zhu_ingestible_2023}, it uses a centimeter-scale battery for power and lacks wireless charging. Both wireless power transfer and communication are necessary for chronic use of these devices.

\begin{figure}[!t]
\centering
\includegraphics[width=3.49in]{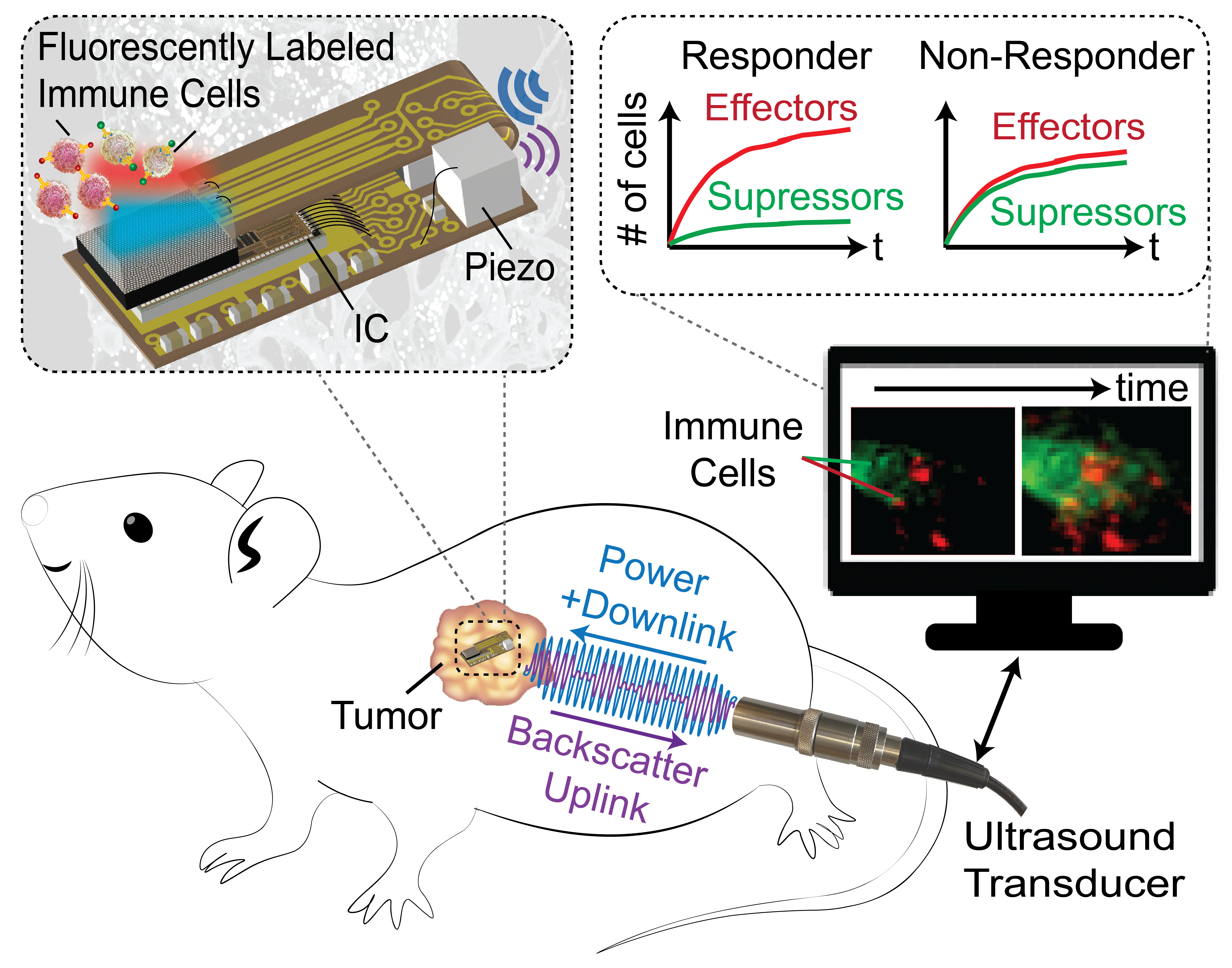}
\caption{Concept of a fully wireless, multicolor, implantable imager for real-time monitoring of immune response.}
\label{intro}
\end{figure}

Here we present a fully wireless, miniaturized fluorescence image sensor capable of three-color fluorescence imaging, aiming to enable real-time, chronic monitoring of cellular interactions deep in the body (Fig. \ref{intro}). Wired connections and batteries are eliminated by power harvesting and bi-directional communication through ultrasound (US). Among wireless power transfer modalities such as near-field inductive coupling, RF, and optical, US offers low loss in tissue (0.5--1 dB/MHz/cm \cite{chen_acoustic_2022}), a high Food and Drug Administration (FDA) regulatory limit for power density (720 mW/cm\textsuperscript{2}), and a short wavelength ({\textapprox}3--4 mm in the PZT material at 1 MHz) enabling power transfer to millimeter-scale implants at centimeter-scale depths \cite{singer_wireless_2021, basaeri_review_2016}. 

While significant progress toward a wireless fluorescence imaging system using US is presented in our prior work \cite{rabbani_3640_2022, rabbani_towards_2024, rabbani_towards_2021}, this system has several limitations. It incorporates a large (0.18 cm\textsuperscript{3}) {\textapprox}1 mF off-chip capacitor for energy storage. It only operates at 2 cm depth, constraining its application to superficial tumors while exceeding FDA US safety limits by 26\% due to high acoustic power requirements. Moreover, the sensor only images a single fluorescent channel, lacking the necessary hardware for multicolor imaging such as a wirelessly programmable laser driver to control multiple excitation lasers and a multi-bandpass optical filter. Additionally, due to in-pixel leakage during readout, the sensitivity of the imager when operating wirelessly is limited to high concentrations of fluorophores, rendering it insufficient for imaging biologically relevant samples. 

This work demonstrates a new system with significant improvements in performance and size, specifically designed for multicolor imaging. Our new system shows fully wireless operation at 5 cm depth in oil, requiring 221 mW/cm\textsuperscript{2} US power flux density (31\% of FDA limits) for power harvesting and transmitting data with a bit error rate (BER) less than 10\textsuperscript{-6} through US backscatter. It powers three different-wavelength laser diodes programmed through US downlink and incorporates a multi-bandpass optical frontend expanding on the design in \cite{roschelle_multicolor_2024} to enable three-color fluorescence imaging. Moreover, we illustrate the application of our sensor in assessing response to cancer immunotherapy through multicolor fluorescence imaging of both effector and suppressor immune cells in \textit{ex vivo} mice tumor samples with and without immunotherapy. Finally, a proof-of-concept mechanical assembly demonstrates a small form factor of 0.09 cm\textsuperscript{3}.

This article further explains and expands on the work presented in \cite{rabbani_173_2024} and is organized as follows. Section II discusses the components and design specifications for a fully wireless, multicolor fluorescence imager. We describe the design and implementation of our system in Section III. Section IV presents system-level measurement results. We illustrate the application of our sensor with \textit{ex vivo} imaging results in Section V. Finally, Section VI includes a comparison with the state of the art and the conclusion.
\section{System Overview}
\begin{figure}[!t]
\centering
\includegraphics[width=3.49in]{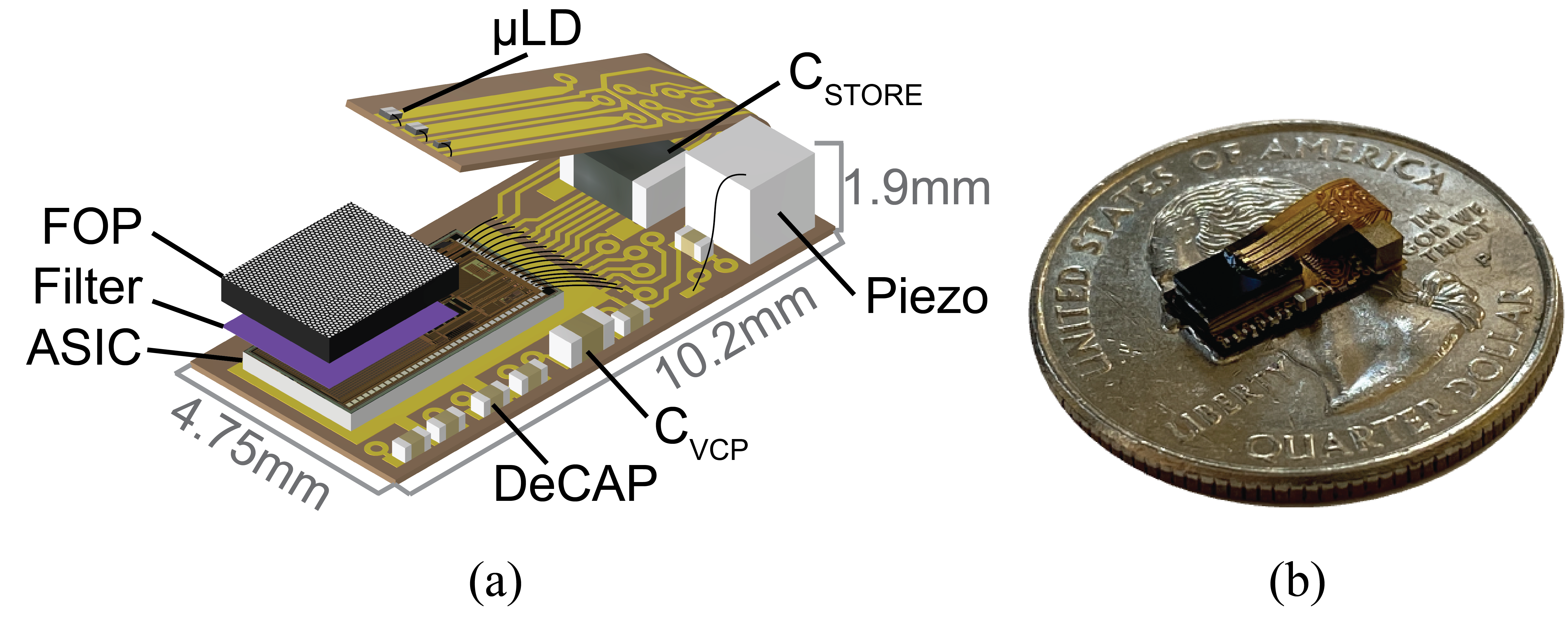}
\caption{(a) To-scale diagram of the full system. (b) Mechanical assembly.}
\label{system overview}
\end{figure}

Fig. \ref{system overview} shows a diagram and mechanical assembly of the full system on a flex PCB with all external components. The system consists of: 1) micro-laser diodes ({\textmu}LDs) for in-situ illumination; 2) an optical frontend comprising of a fiber optic plate and a multi-bandpass interference filter for lens-less multicolor fluorescence imaging; 3) a piezoceramic as the US transceiver; 4) off-chip capacitors for energy storage; and 5) an ASIC to integrate all of this functionality. In this section, we will describe the design of the components in the system and derive design requirements for the ASIC.
\subsection{Multicolor Fluorescence Imaging}
\begin{figure}[!t]
\centering
\includegraphics[width=3.49in]{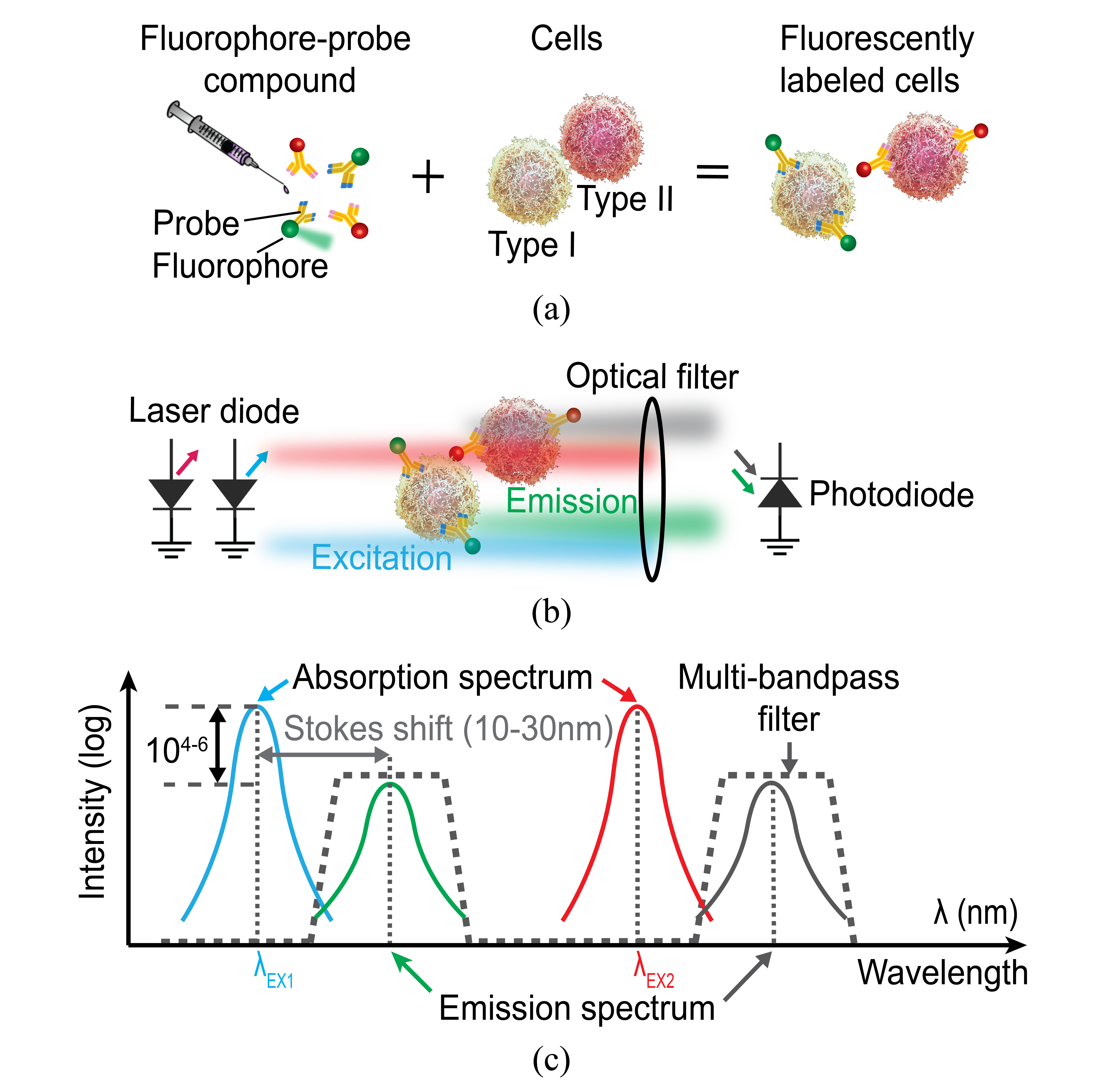}
\caption{Multicolor fluorescence imaging. (a) Each cell type is labeled with a different color fluorescent probe. (b,c) Fluorophores are excited near the absorption peak and emit light at a slightly longer wavelength. A multi-bandpass filter passes emissions while blocking excitation.}
\label{multicolor fluorescence}
\end{figure}

Fig. \ref{multicolor fluorescence} illustrates the principle of multicolor fluorescence imaging. The fluorophores are first conjugated to a probe (Fig. \ref{multicolor fluorescence}(a)), such as an antibody, targeted toward a cell type of interest \cite{lichtman_fluorescence_2005}. For \textit{in vivo} imaging, the conjugated probe can be administered systemically through intravenous injection, binding only to targeted cells. Many organic fluorophores have low toxicity at doses relevant for imaging \cite{alford_toxicity_2009} and a number of fluorescent probes are FDA-approved or in clinical trials, including some using Fluorescein (FAM) and Cyanine5 (Cy5) \cite{barth_fluorescence_2020}, the fluorophores in our \textit{ex vivo} studies. Once injected, the half-life of antibody-based probes is days to weeks \cite{freise_vivo_2015} and free-floating unbound probes are cleared through the liver and kidneys in 1–7 days \cite{mieog_fundamentals_2022}. 

After labeling the cells, the fluorophores are excited near their absorption peak ({$\lambda$}\textsubscript{EX}) and emit light at a slightly longer wavelength with a peak at {$\lambda$}\textsubscript{EM} (Fig. \ref{multicolor fluorescence}(b) and (c)). For organic fluorophores, the difference between the absorption and emission peaks, or Stokes shift, is 10–30 nm (26 nm for FAM and 18 nm for Cy5). Moreover, due to the small absorption cross-section of the fluorophores relative to the illuminated field of view (FoV), the excitation light is often 4 to 6 orders of magnitude stronger than the emission light. Thus, in order to detect the weak fluorescence signal, an optical filter with an optical density (OD) $\ge$ 6 is required to attenuate out-of-band excitation light that would otherwise saturate the sensor. Avoiding a filter altogether through time-gated imaging \cite{moazeni_mechanically_2021, choi_512-pixel_2019, najafiaghdam_optics-free_2022}---where excitation and imaging are separated in the time domain---leads to inadequate excitation rejection and low signal intensities with typical organic fluorophores, which have fluorescence lifetimes less than 10 ns \cite{berezin_fluorescence_2010}. Moreover, background subtraction in the electrical domain \cite{aghlmand_65-nm_2023} adds additional noise sources and is challenging \textit{in vivo} as the excitation background is dependent on tissue scattering.

For multicolor imaging, a variety of organic fluorophores are available with absorption and emission wavelengths spanning the visible and NIR spectrum \cite{haugland_handbook_1992}. Their narrow absorption and emission spectra allow for multiplexed imaging using a monochrome sensor by taking a separate image at each excitation wavelength. Therefore, multicolor fluorescence imaging requires multiple excitation sources and a multi-bandpass filter to block all excitation wavelengths while passing fluorescence emissions. 
\subsection{Light Sources}
For fluorescence excitation, we use {\textmu}LDs with wavelengths of 650 nm (250×300×100 {\textmu}m\textsuperscript{3}, CHIP-650-P5, Roithner LaserTechnik GmbH) and 455 nm (120×300×90 {\textmu}m\textsuperscript{3}, LS0512HBE1, Light Avenue). A third 785 nm laser diode (L785P5, ThorLabs) in a TO-can package is used for proof-of-principle three-color fluorescence imaging and will be replaced by a {\textmu}LD in the future. Laser diodes are chosen instead of LEDs which have broader spectral bandwidths that can overlap with fluorescence emissions. These out-of-band emissions necessitate excitation filters on the LEDs that complicate sensor design and waste optical power output \cite{azmer_miniaturized_2021}. 

\begin{figure}[!t]
\centering
\includegraphics[width=3.49in]{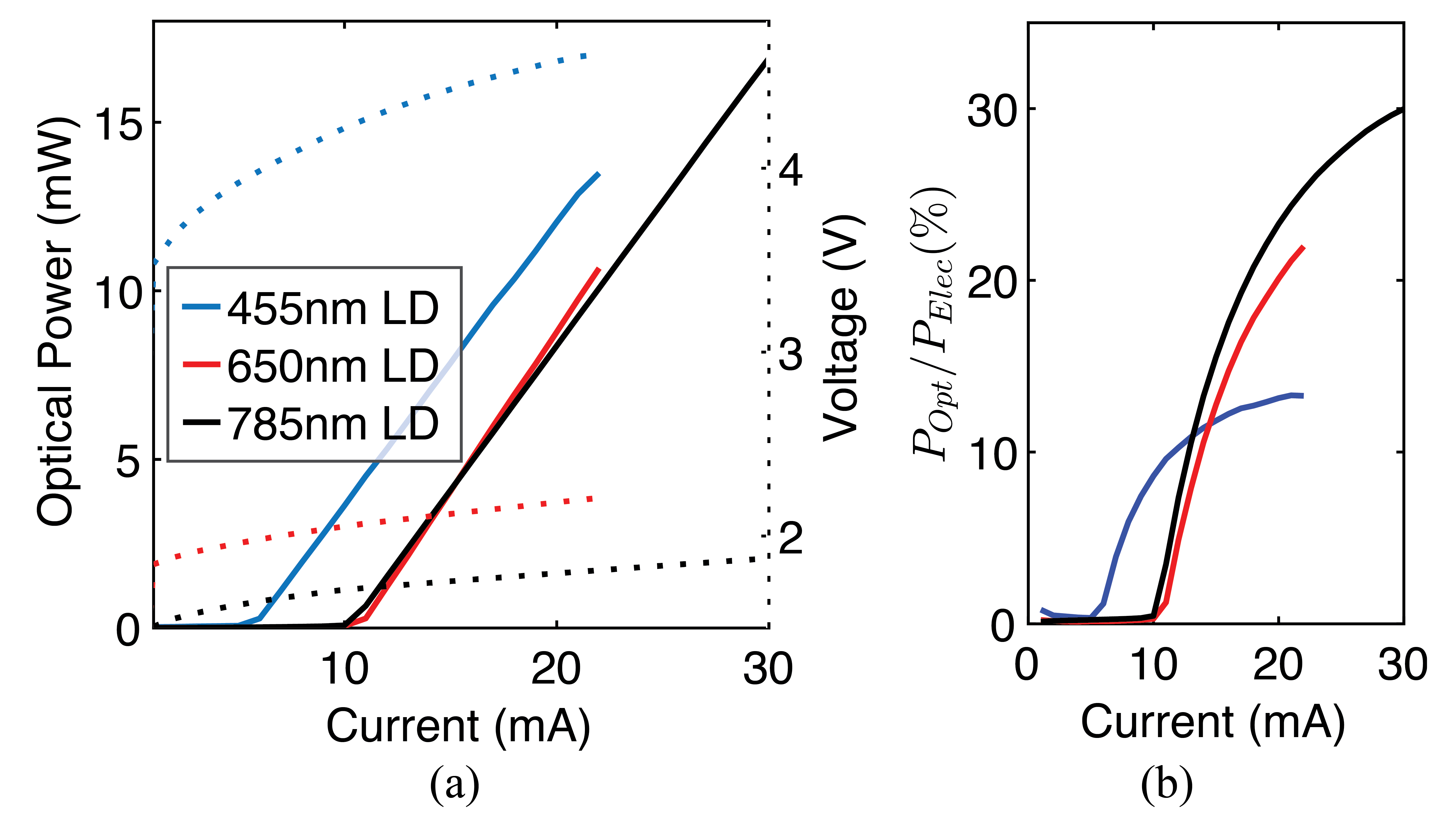}
\caption{Measured laser diode (a) PIV curves and (b) wall-plug efficiencies.}
\label{laser char}
\end{figure}

Fig. \ref{laser char}(a) and (b) show the measured power-current-voltage (PIV) curves for all three lasers and their calculated wall-plug efficiencies ($P_{Optical}/P_{Electrical}$), respectively. The lasers have different forward voltages: {\textapprox}2 V for the 650 nm and 785 nm lasers and {\textapprox}4.5 V for the 455 nm laser. Because of their several-mA threshold currents, the lasers operate most efficiently near their maximum current ratings. These characteristics motivate the design of a laser driver with programmable current that is tolerant of a wide range of forward voltages. 
\subsection{Optical Frontend Design}
The optical frontend design builds on our prior work \cite{roschelle_multicolor_2024} and consists of a multi-bandpass interference filter and a low-numerical-aperture fiber optic plate (FOP). Interference filters offer more-ideal filter characteristics than absorption filters \cite{papageorgiou_angle-insensitive_2018} or CMOS metal filters \cite{aghlmand_65-nm_2023, zhu_ingestible_2023, hong_fully_2017}, which do not allow for optimal excitation and imaging of organic fluorophores due to their gradual cutoff transitions, weak out-of-band attenuation, and significant passband losses. Hybrid filters combining interference and absorption filters \cite{moazeni_mechanically_2021, rustami_needle-type_2020, sasagawa_highly_2018} retain the poor passband characteristics of absorption filters. Another major advantage of interference filters is their ability to support multiple passbands across the visible and NIR spectra for multicolor imaging. In contrast, demonstrated dual-color fluorescence sensors with absorption or CMOS filters rely on dedicated pixels for each color \cite{aghlmand_65-nm_2023, hee_lens-free_2019, taal_toward_2022, kulmala_lensless_2022}, reducing the sensor sensitivity and resolution. 

However, interference filters are sensitive to angle of incidence (AOI) \cite{dandin_optical_2007}. At increasing AOIs, the filter passbands shift towards shorter wavelengths, eventually transmitting the excitation light. This property is problematic for lensless imaging where the AOI is not precisely controlled and the excitation light is often angled between the sensor and the tissue above it. To mitigate this effect, the FOP acts as an angle filter, blocking off-axis excitation light that would otherwise pass through the filter. The FOP also improves resolution by eliminating divergent fluorescent emissions that contribute to blur, albeit at the cost of reducing the overall collected signal.

\begin{figure}[!t]
\centering
\includegraphics[width=3.49in]{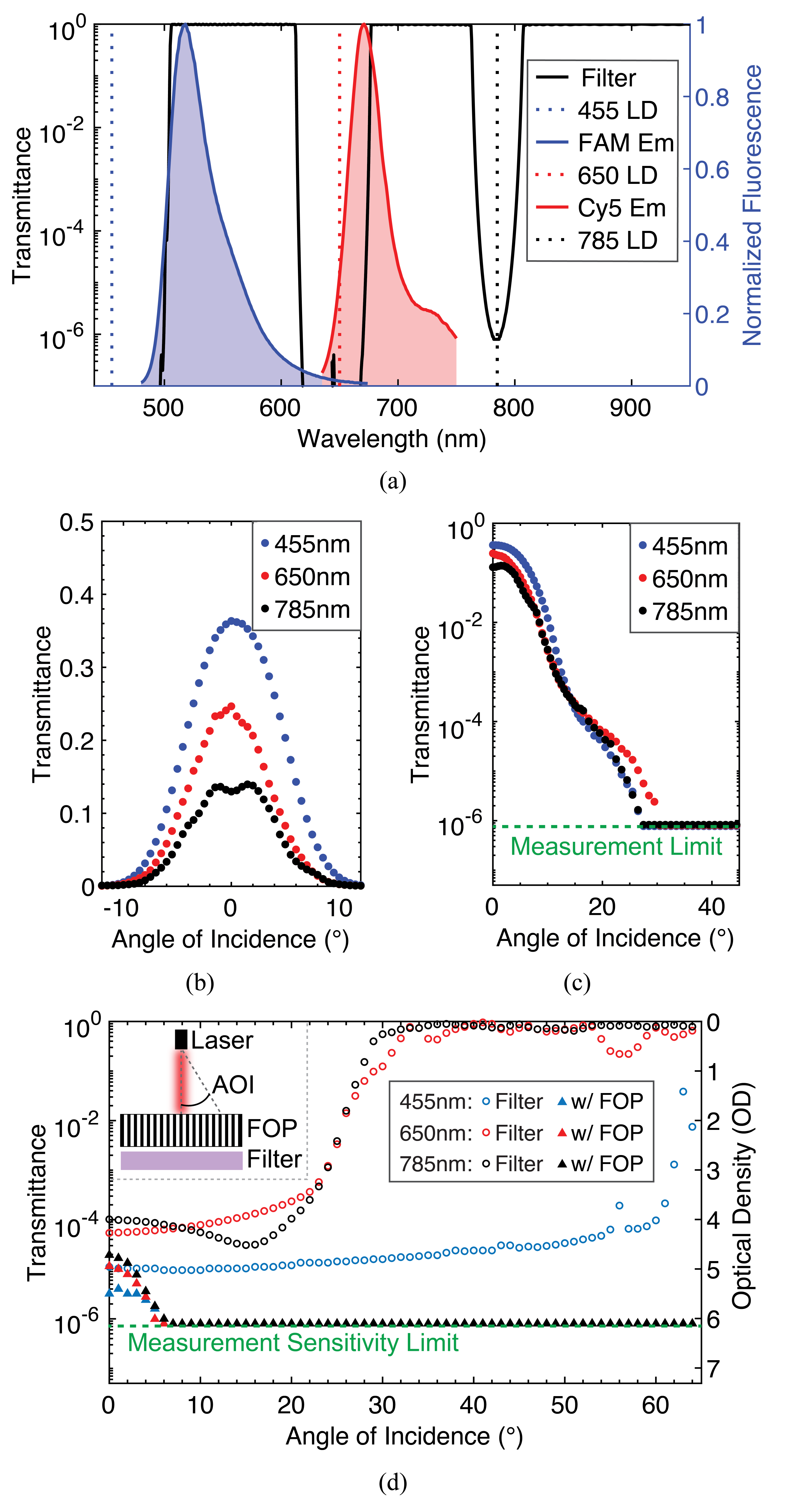}
\caption{(a) Normal incidence transmittance spectra of the multi-bandpass interference filter. (b,c) Measured transmittance through the FOP across AOIs. (d) Angular transmittance of the filter with and without the FOP measured at the excitation laser wavelengths.}
\label{optical frontend char}
\end{figure}

Here, we expand the dual-bandpass design in \cite{roschelle_multicolor_2024} to three-color fluorescence imaging with a new interference filter. Fig. \ref{optical frontend char}(a) shows the normal incidence (AOI=0°) transmittance spectra of the filter (ZET488/647/780+800lpm, Chroma Technologies Corp) which has three passbands with greater than 93\% average transmittance. The first two bands pass the emissions of FAM and Cy5, the fluorophores used in our \textit{ex vivo} imaging studies. The 800 nm band, added in this work, provides another fluorescence channel in the NIR-I window (700–900 nm), a preferred region for \textit{in vivo} imaging where tissue scattering, absorption, and autofluorescence are minimal compared to the visible spectrum (400–700 nm) \cite{frangioni_vivo_2003, ji_near-infrared_2020}. At normal incidence, the filter provides sufficient blocking of the lasers: more than 6 OD attenuation at both 450 nm and 650 nm as well as more than 5 OD attenuation at 785 nm.

The 500 {\textmu}m-thick FOP (LNP121011, Shenzhen Laser, LTD) consists of a matrix of 10 {\textmu}m optical fibers embedded in black, absorptive glass. It has a normal incidence transmittance of 35\% and a full-width at half maximum (FWHM) of 10° at 455 nm, which both reduce at longer wavelengths as shown in Fig. \ref{optical frontend char}(b). The angular transmittance measurements in Fig. \ref{optical frontend char}(c) show that beyond an AOI of 35° the FOP provides more than 6 OD attenuation of all three lasers. 

Fig. \ref{optical frontend char}(d) shows the transmittance through the filter with and without the FOP across different AOIs measured at the excitation wavelengths using collimated, fiber-coupled lasers. The filter attenuation at AOI=0° is different from that in Fig. \ref{optical frontend char}(a) due to out-of-band emissions from the lasers. While the filter blocks the excitation lasers near 0°, the laser transmittance rapidly increases beyond AOIs of 20° for 650 nm and 785 nm and 60° for 455 nm. However, with the FOP, the optical frontend provides more than 6 OD of attenuation of all excitation lasers at AOIs greater than 5°. The maximum measured attenuation is limited by the sensitivity of the power meter (PM100D with S120C Photodiode, Thorlabs) used for this measurement.

For fabrication, the interference filter is directly deposited on the FOP, resulting in a total thickness of approximately 510 {\textmu}m. The optical frontend is fixed to the chip using optically transparent epoxy (SYLGARD 184, Dow Chemicals). The filter is placed in between the chip and the FOP to ensure that it blocks any excitation light scattered through the FOP \cite{roschelle_multicolor_2024}. 

\subsection{Ultrasound Link}
\begin{figure}[!t]
\centering
\includegraphics[width=3.49in]{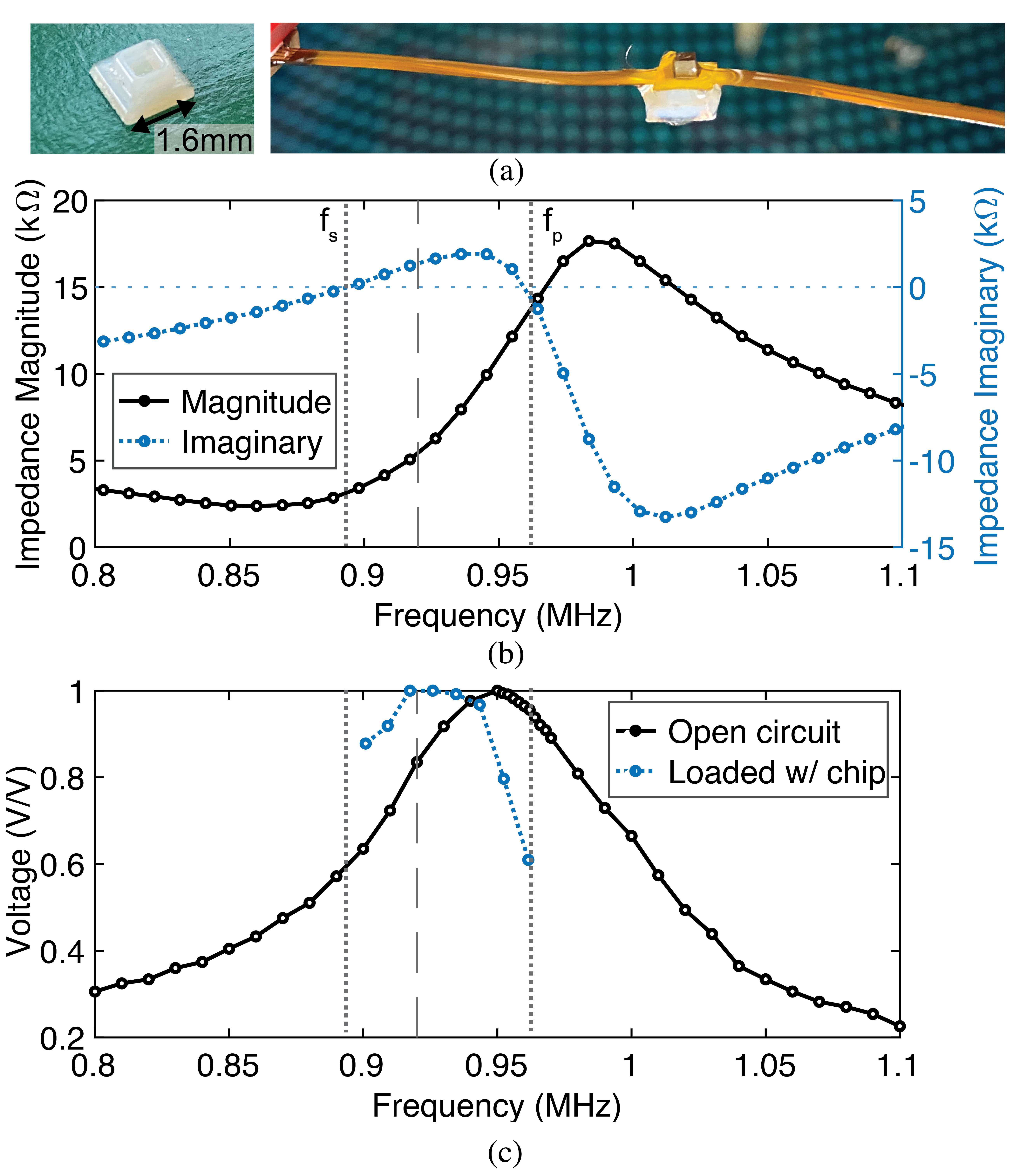}
\caption{(a) Piezo assembly with the air gap. (b) Measured electrical impedance of the piezo across frequency. (c) Measured harvested voltage across frequency with the piezo in open circuit condition and loaded by the chip.}
\label{piezo char}
\end{figure}

We use a 1.5×1.5×1.5 mm\textsuperscript{3} piezoceramic (lead zirconate titanate) as the US transceiver for wireless power transfer and bi-directional communication. The thickness of the piezo is directly proportional to the harvested voltage and inversely proportional to the operation frequency \cite{singer_wireless_2021}. Therefore, we chose a thickness of 1.5 mm to balance minimizing the overall size of the piezo with the need for harvesting a high enough voltage to drive the lasers while operating at a lower frequency with less tissue attenuation. An aspect ratio of one is selected as a compromise between volumetric efficiency and backscattering amplitude, as outlined in \cite{ghanbari_optimizing_2020}.
The piezo is mounted on a flex PCB for testing (Fig. \ref{piezo char}(a)). On the backside of the piezo, an air gap is created by covering a through-hole via with a 3D-printed lid. The air gap reduces the acoustic impedance of the backside medium from 1.34 MRayl in canola oil to {\textapprox}0 MRayl in air, decreasing the electrical impedance of the piezo to improve the power transfer efficiency \cite{sonmezoglu_method_2021}. 

Fig. \ref{piezo char}(b) shows the impedance spectrum of the piezo measured within canola oil. Canola oil has 0.075 dB/cm acoustic attenuation at 920 kHz and 1.34 MRayl acoustic impedance \cite{gladwell_ultrasonic_1985} similar to the impedance (1.4–1.67 MRayl) of tissue \cite{chen_acoustic_2022}. The series and parallel resonance frequencies of the piezo occur at, f\textsubscript{S}=894 kHz and  f\textsubscript{P}=960 kHz, respectively. Fig. \ref{piezo char}(c) shows the normalized harvested voltage across frequency when the piezo is open circuit condition and when it is loaded with the chip (see section IV.A for the setup). While operating near f\textsubscript{S} minimizes the impedance, the open circuit voltage is maximized near f\textsubscript{P}. Therefore, the maximum harvested voltage with the chip occurs between f\textsubscript{S} and f\textsubscript{P} at 920 kHz. 
\subsection{System Design Considerations}
To derive the required harvested energy per image for sizing the storage capacitor, we estimate the signal detected by a pixel from Cy5-labeled CD8+ T-cells, a type of immune cell imaged in our \textit{ex vivo} studies. The total emitted optical power, $P_{CELLS}$, from $C$ fluorescently labeled cells as a function of the input excitation  intensity, $I_{IN}$, is given by
\begin{equation}
\label{PCELLS}
P_{CELLS}=C\cdot N_{FL} \cdot \sigma \cdot QY \cdot I_{IN}.
\end{equation}
$N_{FL}$ is the number of fluorophores bound to each cell. Typically, between 0.5–2.1×10\textsuperscript{6} CD8+ antibodies bind to a single CD8+ T-cell \cite{siiman_cell_2000} with each antibody containing 2–8 fluorophores \cite{vira_fluorescent-labeled_2010}. $\sigma$ and $QY$ are the absorption cross-section and quantum yield of the fluorophore, respectively (9.55×10\textsuperscript{-16} cm\textsuperscript{2} and 20\% for Cy5 \cite{aat_bioquest_extinction_2024}). We assume that a single pixel (with 55 {\textmu}m pitch in our design) subtends a FoV containing $C=100$ T-cells, considering that a T-cell is 5–10 {\textmu}m in diameter \cite{jiang_mri_2020}. Assuming that the 650 nm {\textmu}LD uniformly illuminates the FoV of our sensor (2×2.2 mm\textsuperscript{2}) and outputs 10 mW of optical power at I\textsubscript{LD}=20 mA bias (see Fig. 4), $I_{IN}$ is approximately 223 mW/cm\textsuperscript{2}. Therefore, the estimated total fluorescence signal is 20 nW. This signal can be converted to the expected photodiode current, $I_{PH}$, according to
\begin{equation}
\label{IPH}
I_{PH}=P_{CELLS}\cdot \frac{A_{PIXEL}}{4 \pi z_{DIST}^{2}}\cdot (1- L_{FOP}) \cdot R.
\end{equation}
This equation accounts for both the spreading loss over the $z_{DIST}\approx 500$ {\textmu}m distance to the pixel with area, $A_{PIXEL}$ (44×44 {\textmu}m\textsuperscript{2} in our design) and the insertion loss of the FOP, $L_{FOP}$ ({\textapprox}75\% at 650 nm). Given that the pixel has a responsivity, $R$, of 0.21 A/W at 650 nm, we expect $I_{PH}$ on the order of 6.3 fA.

In the capacitive trans-impedance amplifier (CTIA)-based pixel architecture reused from \cite{papageorgiou_chip-scale_2020} the photocurrent is sensed by integrating it on a capacitor,  $C_{INT}$, during the exposure time, $T_{EXP}$, resulting in a pixel output voltage of
\begin{equation}
\label{CTIA}
V_{PIXEL}=\frac{I_{PH}\cdot T_{EXP}}{C_{INT}}.
\end{equation}
Sensing the fluorescence signal relies on $V_{PIXEL}$ exceeding the noise floor, characterized by the signal-to-noise ratio (SNR). Generally, SNR can be improved by increasing the total imaging time either through a longer exposure time, $T_{EXP}$, or by averaging multiple images. Following the derivation in [32], the SNR at the output of a CTIA-based pixel when averaging $n$ images with an exposure time of $\frac{T_{EXP}}{n}$ is given by
\begin{equation}
\label{SNR}
SNR(n\cdot \frac{T_{EXP}}{n})=\frac{signal}{noise}=\frac{\frac{I_{PH}T_{EXP}}{C_{INT}}}{\sqrt{\frac{T_{EXP}}{C_{INT}^2}}2q_{e}i_{D}+n \overline{v_{NR}^2}}.
\end{equation}
This equation enables study of the SNR tradeoff between (1) taking a single exposure of  ($n$=1) and (2) averaging $n$ images with exposures of $\frac{T_{EXP}}{n}$. The noise has two components: readout noise, $\overline{v_{NR}^2}$, and shot noise from the photocurrent and dark current, $i_{D}=I_{PH}+I_{DARK}$. $q_{e}$ is the charge of an electron. The factor of $n$ only appears in the readout noise term. Therefore, if shot noise is the dominant source of noise, for small $n$, both (1) and (2) result in the same SNR. However, with increasing  and lower exposure time per frame, readout noise dominates the overall noise of the averaged image, necessitating a greater number of averages to maintain the same SNR as a single exposure. 

Using the estimated $I_{PH}$ and the measured noise values reported in Section IV, we calculate that without averaging, a $T_{EXP}$ of 98 ms is required to achieve an SNR of 20 dB (10×). This result corresponds to a minimum required energy ( $I_{LD}\cdot V_{LD}\cdot T_{EXP}$) of 4.16 mJ per image. 

Delivering $I_{LD}$=20 mA from the incident US signal, given a piezo impedance of 5.4 k$\Omega$ at 920 kHz, requires an open circuit voltage of at least 108 V, which is not practical within FDA limits. Therefore, harvested energy must first be stored on a capacitor to later supply the lasers when taking an image. The size of the storage capacitor, $C_{STORE}$, is determined by $C_{STORE}=\frac{I_{LD}T_{EXP}}{\Delta V_{CSTORE}}$ in order to supply $I_{LD}$ for the duration of $T_{EXP}$. $\Delta V_{CSTORE}$ is the voltage drop on the capacitor during $T_{EXP}$. Maximizing $\Delta V_{CSTORE}$ results in a smaller capacitor size, but is limited by the maximum harvested voltage and the minimum supply requirements for operating the chip or laser. Assuming $\Delta V_{CSTORE}$=3 V, results in a capacitor size of 650 {\textmu}F. Capacitors of this size are large physical components, increasing implant volume as in \cite{rabbani_towards_2024}. Therefore, the capacitor size can be minimized by reducing the required energy per image through the averaging strategy discussed previously. 

\begin{figure}[!t]
\centering
\includegraphics[width=3.49in]{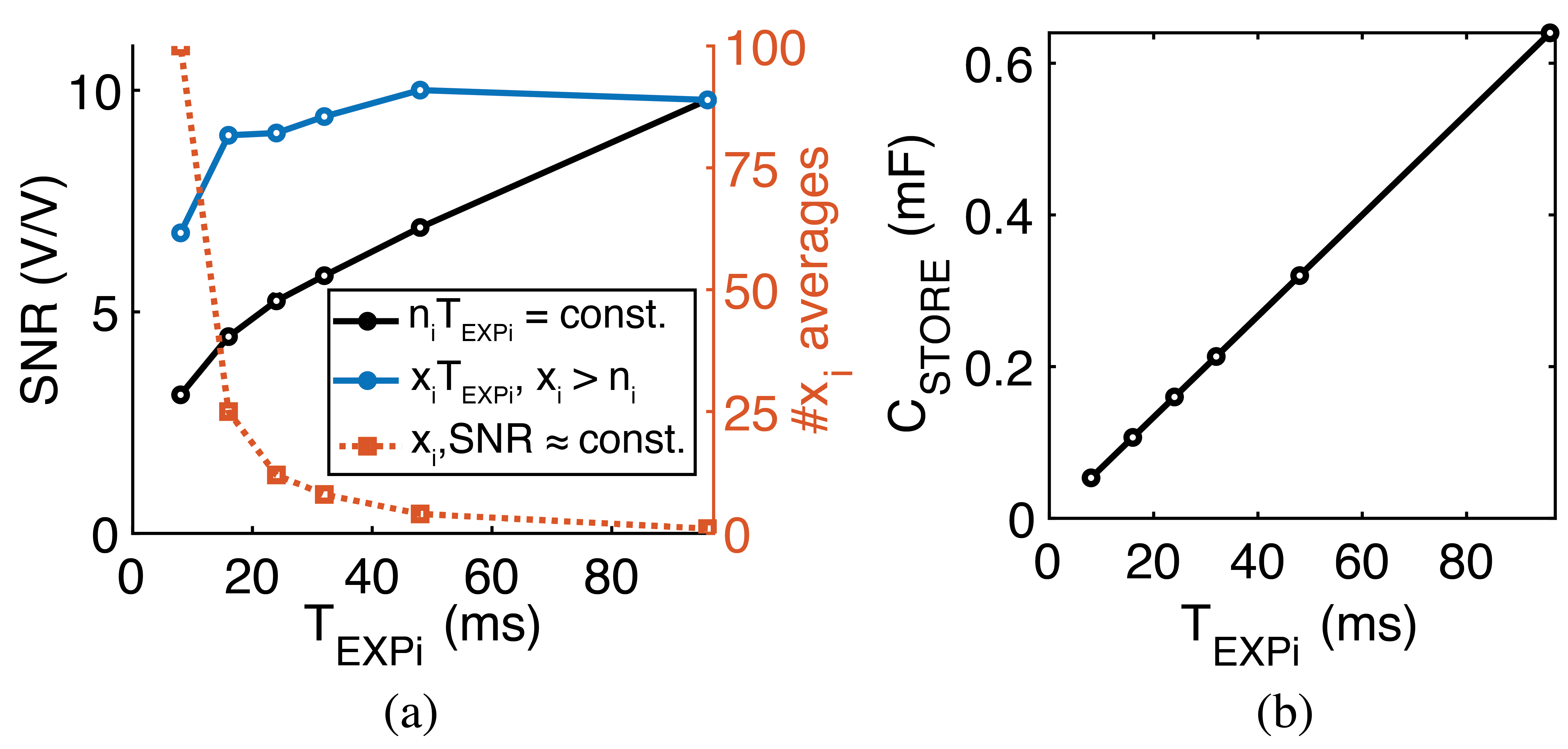}
\caption{(a) In black, the SNR of a pixel with the estimated photocurrent and measured dark noise (see Fig. \ref{pixel char}(c)) across different exposure times (T\textsubscript{EXP,i}) and averaging n\textsubscript{i} images such that total imaging time $T_{EXPi}n_{i}=96$ ms remains constant. In blue, the SNR after $x_i$ averages (orange) required to maintain 90\% of the SNR at $T_{EXPi}= 96 ms$. (b) Capacitor size vs. exposure time.}
\label{system design}
\end{figure}

Fig. \ref{system design}(a) compares the SNR of a pixel with different levels of averaging. The signal is the estimated photocurrent from the above analysis (6.3 fA) and the noise is measured with the sensor from dark images (see Fig. \ref{pixel char}(c)). Each data point on the black curve represents an exposure time of T\textsubscript{EXPi} and a number of averages n\textsubscript{i} such that the total exposure time, n\textsubscript{i}T\textsubscript{EXPi}=96 ms stays constant. As T\textsubscript{EXPi }decreases (and n\textsubscript{i} increases), readout noise dominates the pixel output noise (because shot noise decreases with lower T\textsubscript{EXPi}), requiring additional averages to achieve the same SNR of a single exposure. The orange curve in Fig. \ref{system design}(a) shows the increased number of averages, x\textsubscript{i} > n\textsubscript{i}, required to reach an SNR (shown in blue) within 90\% of the initial SNR for T\textsubscript{EXP}=96 ms. Therefore, using averaging to decrease exposure time for individual frames increases the overall imaging time to greater than 96 ms. As shown in Fig. \ref{system design}(b), the capacitor size decreases linearly with lower T\textsubscript{EXPi} ranging from 640 {\textmu}F for T\textsubscript{EXPi}=96 ms to 50 {\textmu}F for T\textsubscript{EXPi}=8 ms. Charging such a capacitor through US takes several seconds to minutes, dominating the frame time (see Section IV.B). Thus, for small exposure times, the additional required averages can significantly increase the total imaging time. The total imaging time must be less than several minutes to capture the motion of immune cells, which have mean velocities of 10 {\textmu}m/min in the tumor microenvironment \cite{dupre_t_2015}.

Following these guidelines, we chose an 0805 100 {\textmu}F tantalum capacitor for C\textsubscript{STORE} with a size of 2×1.25×0.9 mm\textsuperscript{3} (0.002 cm\textsuperscript{3}). This capacitor can supply 20 mA of laser current for T\textsubscript{EXP}=16 ms while dropping its voltage by 3 V. Averaging is employed to enhance SNR to levels comparable to those achieved by longer exposure times. We use a tantalum capacitor as opposed to a ceramic capacitor, which can lose up to 40–80\% of its initial capacitance as the DC bias voltage increases and reduces the dielectric permittivity \cite{cen_dc_nodate}.
\section{System Design and Implementation}
\begin{figure*}[!t]
\centering
\includegraphics[width=7.14in]{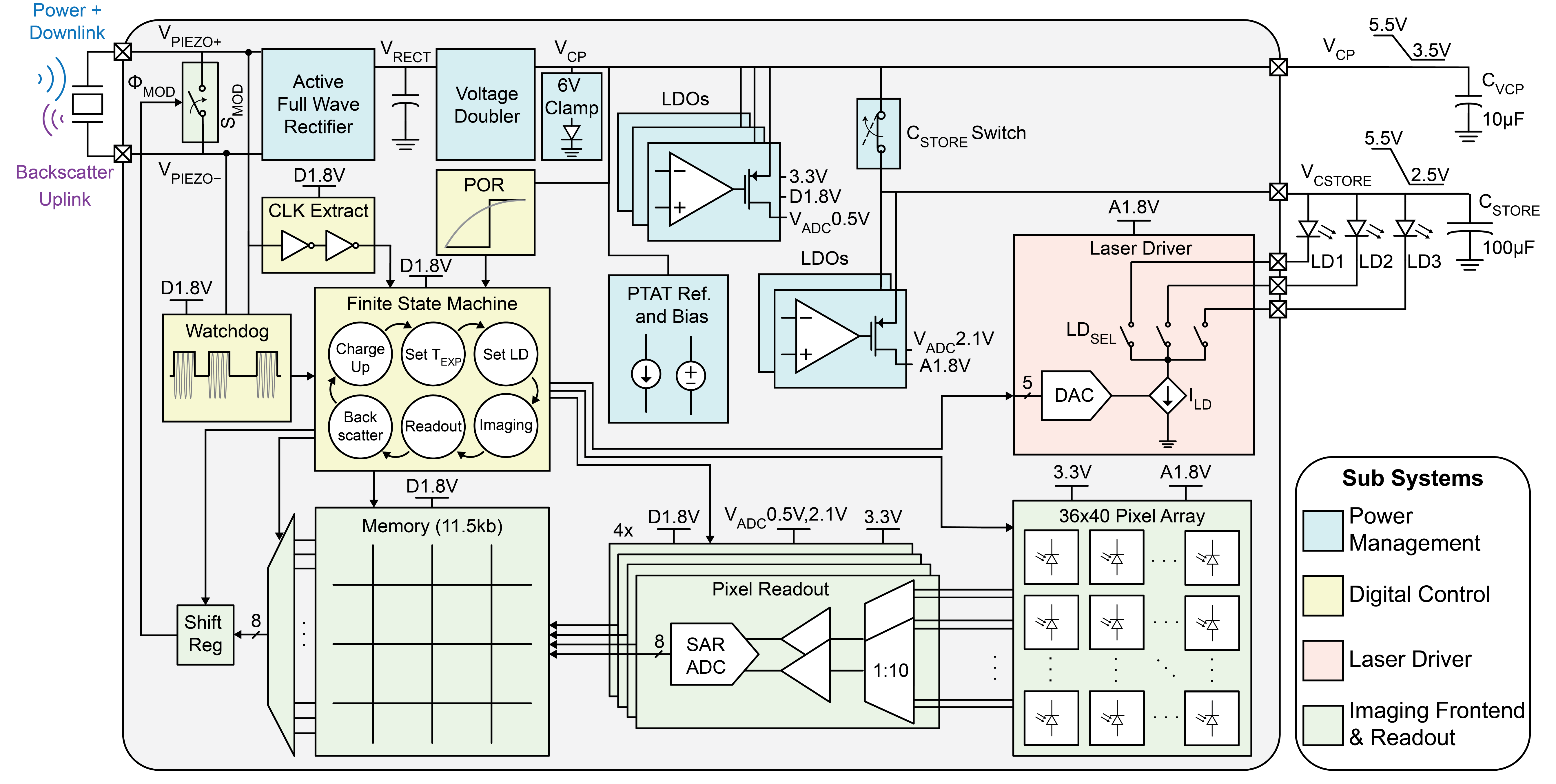}
\caption{System block diagram.}
\label{system diagram}
\end{figure*}

Fig. \ref{system diagram} shows the system block diagram of the ASIC with external connections to the piezo, off-chip storage capacitors, and {\textmu}LDs. The ASIC has 4 main subsystems: (1) power management unit (PMU), (2) digital control, (3) laser driver, and (4) imaging frontend with readout. 

The PMU consists of an active rectifier for AC-DC conversion of the piezo signal and a charge pump for generating an up to 6 V supply for driving the lasers. Harvested energy is stored on two off-chip capacitors, C\textsubscript{VCP}=10 {\textmu}F and C\textsubscript{STORE}=100 {\textmu}F, to separate the power supplies of the lasers from the rest of the sensor throughout its operation. A PTAT develops current and voltage references and several low dropout voltage regulators (LDOs) generate stable DC power supplies for the chip. The sensor is programmed and controlled through a finite state machine (FSM) with 6 states of operation: charging up the storage capacitors (\textit{Charge-Up}); programming the image sensor and laser driver parameters through US downlink (\textit{Set T\textsubscript{EXP}} and \textit{Set LD}); taking an image (\textit{Imaging}); digitizing and storing the image (\textit{Readout}); and wirelessly transmitting the data via US backscatter (\textit{Backscattering}). To take an image, the laser driver, configured during downlink, supplies a {\textmu}LD using energy stored in C\textsubscript{STORE}. The image is captured on a 36×40-pixel array. During \textit{Readout}, the pixel data is digitized by 4 parallel ADCs to be saved in the memory. Finally, image data is transmitted by modulating the reflected amplitude of incident US pulses with the S\textsubscript{MOD} switch. 

The design and operation of the subsystems are described in detail below.
\subsection{Power Management Unit}
\begin{figure}[!t]
\centering
\includegraphics[width=3.49in]{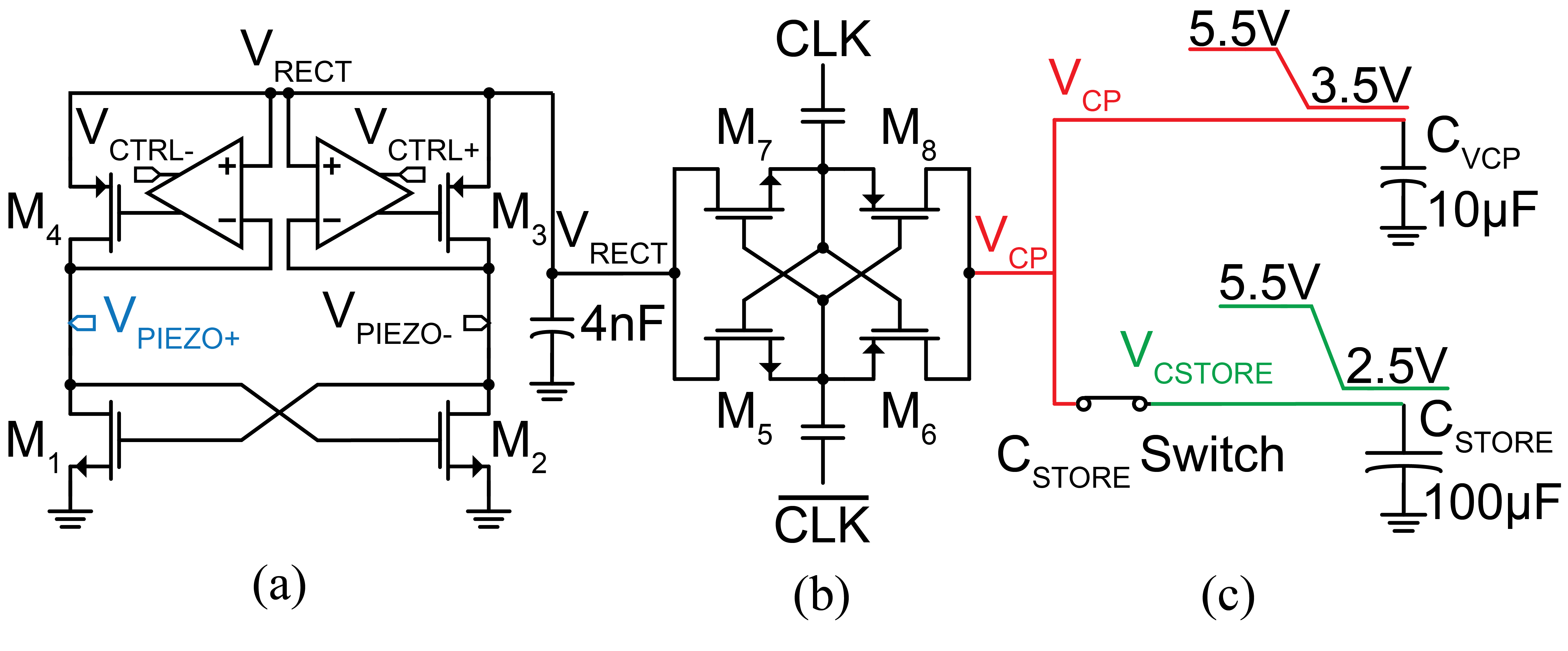}
\caption{PMU schematic consisting of (a) a full-wave active rectifier, (b) a cross-coupled charge pump, and (c) storage capacitors.}
\label{PMU}
\end{figure}

Fig. \ref{PMU} shows the schematic of the active rectifier and charge pump. The active rectifier converts the harvested AC signal on the piezo to a 3 V DC voltage (V\textsubscript{RECT}), which is stabilized by a 4.7 nF off-chip capacitor. V\textsubscript{RECT} is then multiplied by 1.83× to a 5.5 V supply (V\textsubscript{CP}) with the cross-coupled charge pump. The cross-coupled topology is chosen for its high power conversion efficiency for an optimized input range \cite{guler_power_2017}. Compared to a rectifier-only architecture used in \cite{rabbani_towards_2024}, the charge pump reduces the required harvested AC voltage on the piezo (V\textsubscript{PIEZO}) to achieve an output voltage (V\textsubscript{CP}) of 5.5 V by 1.7×, which results in a 3× lower acoustic power density requirement. Acoustic power density is a square function of acoustic pressure, which is linearly proportional to the harvested AC voltage. Therefore, lowering the required harvested piezo voltage reduces the acoustic power density to ensure operation within FDA safety limits. However, with this architecture, the overall charging time increases due to the energy loss from the charge pump.

During \textit{Charge-Up}, C\textsubscript{VCP} and C\textsubscript{STORE} are connected through the C\textsubscript{STORE} switch and are charged through the PMU. C\textsubscript{STORE} stores energy for the lasers and imager array and a smaller C\textsubscript{VCP} stores energy for the readout and digital control. Following manufacturer guidelines, the external US transducer is duty-cycled for reduced average power dissipation to prevent damage to it from overheating while providing enough US power density to achieve sufficient harvested voltage on the sensor. To minimize power consumption during \textit{Charge-Up}, the laser driver, pixel array, readout circuits and memory are switched off. A diode-based voltage clamp prevents charging beyond 6 V to protect the devices from overvoltage.

Five LDOs (Fig. S1) regulate the harvested voltage into stable DC power supplies and are compensated with off-chip 0201 surface mount capacitors (10–200 nF). They generate reference voltages of 0.5 V and 2.1 V for the ADCs, separate 1.8 V power supplies for the digital control and for the pixel array and laser driver biasing, and a 3.3 V supply for the readout. 

\begin{figure}[!t]
\centering
\includegraphics[width=3.49in]{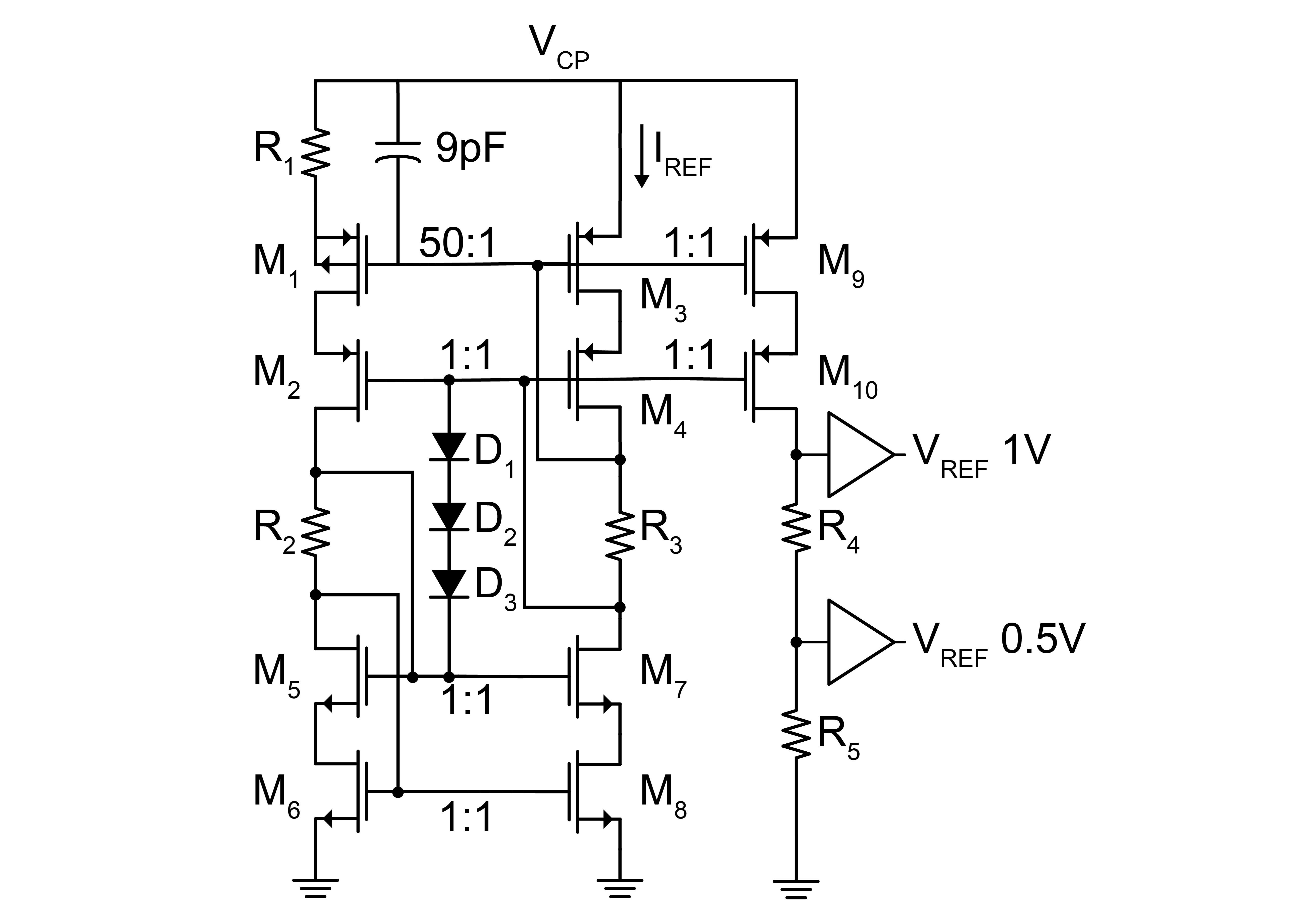}
\caption{PTAT schematic.}
\label{ptat}
\end{figure}

A PTAT circuit generates a 200 nA reference current, I\textsubscript{REF}, and 1 V and 0.5 V references to bias the chip. The PTAT, with schematic shown in Fig. \ref{ptat}, uses a constant-gm topology to minimize the dependence on threshold voltage process variation. A PMOS core (M\textsubscript{1}–M\textsubscript{4}) avoids the body effect as deep N-well transistors were not available in the process. The diode-based start-up circuit (D\textsubscript{1}–D\textsubscript{3}) prevents zero current operation. To ensure that generated references are stable across the large voltage drop on V\textsubscript{CP} from 5.5 to 3.5 V, cascode current mirrors with high output impedance are used throughout the design. The voltage references are buffered and are generated by mirroring I\textsubscript{REF} (M\textsubscript{3}, M\textsubscript{4}, M\textsubscript{9}, M\textsubscript{10}) through resistors R\textsubscript{4} and R\textsubscript{5}. 
\subsection{Digital Control}
\begin{figure*}[!t]
\centering
\includegraphics[width=7.14in]{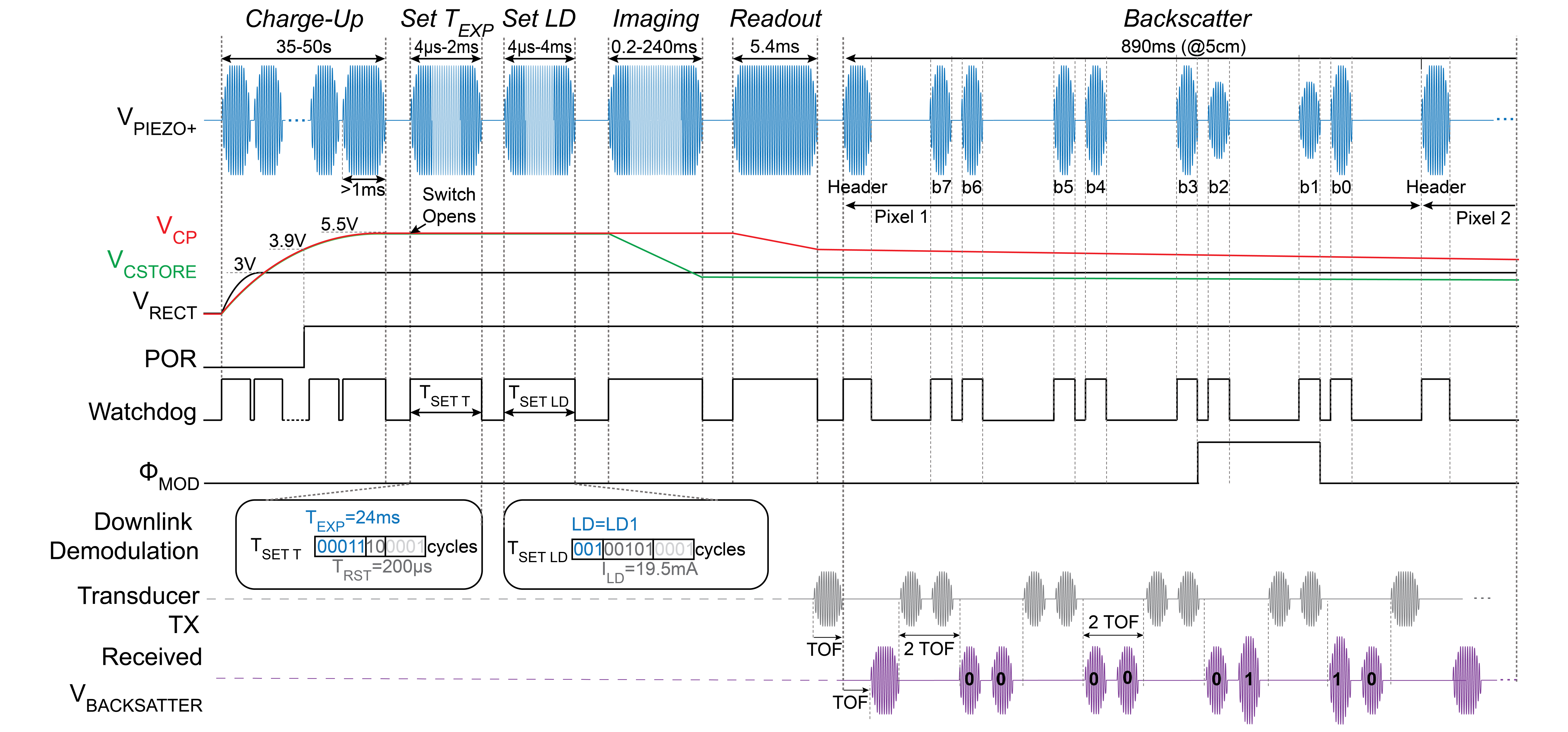}
\caption{System timing diagram.}
\label{timing diagram}
\end{figure*}

The chip operates according to the system timing diagram shown in Fig. \ref{timing diagram}. When V\textsubscript{CP} reaches 3.9 V, ensuring stable operation of the chip, a power-on reset (POR, Fig. S2) circuit initializes the FSM. The FSM is synchronized to the external US transducer by on-off-key modulation of the US envelope, which is demodulated by a watchdog circuit. 

\begin{figure}[!t]
\centering
\includegraphics[width=3.49in]{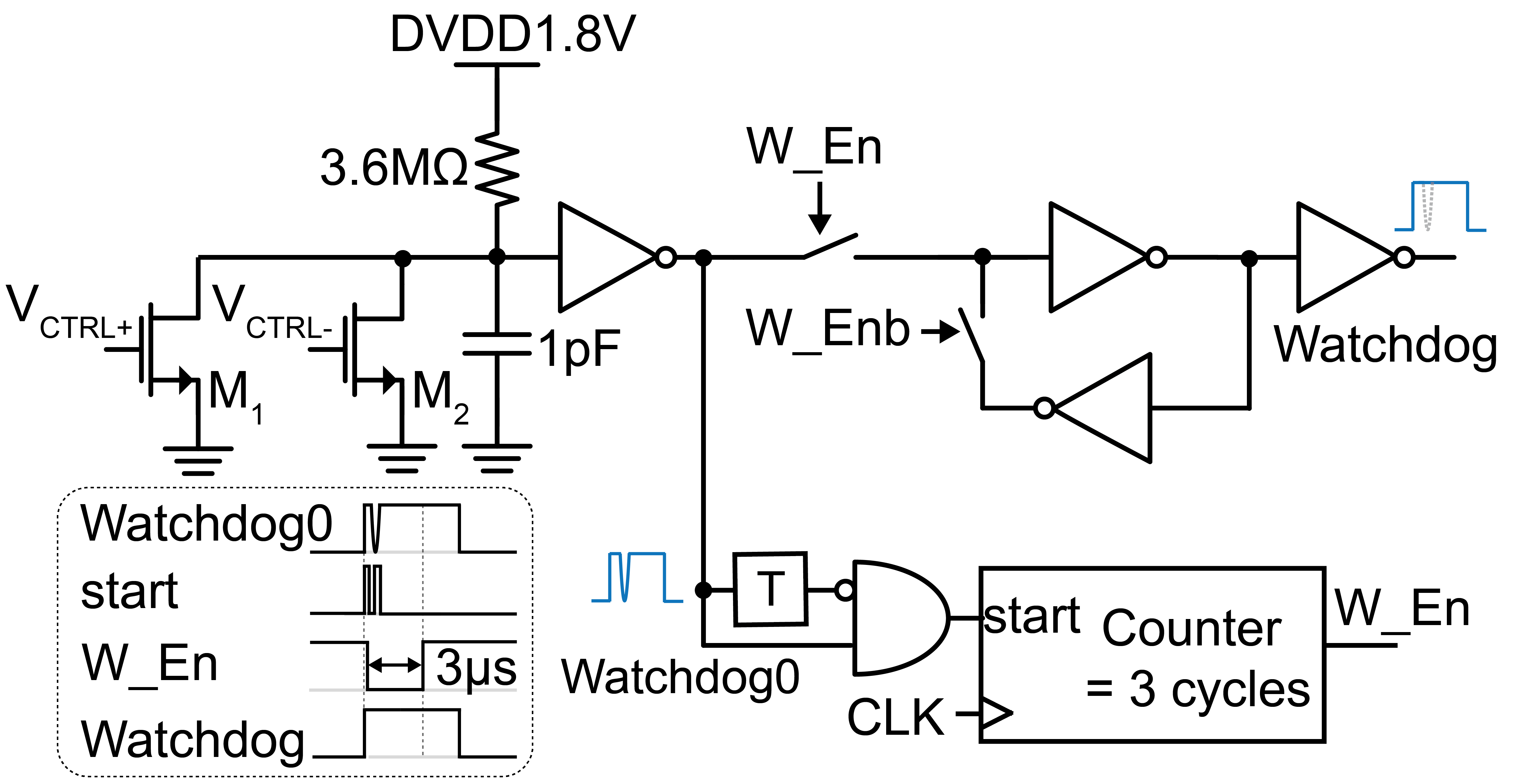}
\caption{Schematic of watchdog circuit with error-tolerant edge detection.}
\label{watchdog}
\end{figure}

The schematic of the watchdog circuit is shown in Fig. \ref{watchdog}. A latched-based control eliminates glitches in detecting the presence of the US pulses within 3 {\textmu}s of the initial rising edge. The unwanted transitions result from insufficient drive strength of the AC inputs to transistors M\textsubscript{1} and M\textsubscript{2} during the gradual ramp-up of the US pulse. 

To relay timing information to the FSM, the clock is extracted from the US carrier frequency (920 kHz). An US pulse longer than 1 ms indicates the end of the \textit{Charge-Up} state. At this moment, the C\textsubscript{STORE} switch is opened to isolate the storage capacitors, allowing V\textsubscript{CSTORE} to drop to a minimum of 2.5 V during \textit{Imaging} while maintaining V\textsubscript{CP} above 3.5 V for the 3.3 V readout circuits. This approach allows for maximum energy usage from C\textsubscript{STORE}, resulting in a 33\% smaller required capacitance assuming a 5.5 V\textit{ Charge-Up }voltage.

After \textit{Charge-Up}, the ASIC is programmed during the \textit{Set T\textsubscript{EXP}} and \textit{Set LD} states. As shown in Fig. \ref{timing diagram}, the transmitted downlink data is decoded through time-to-digital conversion of the US pulse widths. In each state, 4 LSBs are discarded to account for timing variations in the watchdog signal. In \textit{Set T\textsubscript{EXP}}, the exposure time, T\textsubscript{EXP}, is set through the 5 MSBs and is programmable from 0–248 ms with LSB=8 ms. The next 2 bits set the pixel reset time, T\textsubscript{RST}, which can be 100, 200, 500, or 1000 {\textmu}s. In \textit{Set LD}, 3 MSBs set the 1-hot encoded laser channel and the next 5 bits determine the laser current, I\textsubscript{LD}. On the falling edge of the watchdog after \textit{Set LD}, the laser driver and the pixel array bias circuits are turned on to prepare for \textit{Imaging}.

\subsection{Laser Driver}
\begin{figure}[!t]
\centering
\includegraphics[width=3.49in]{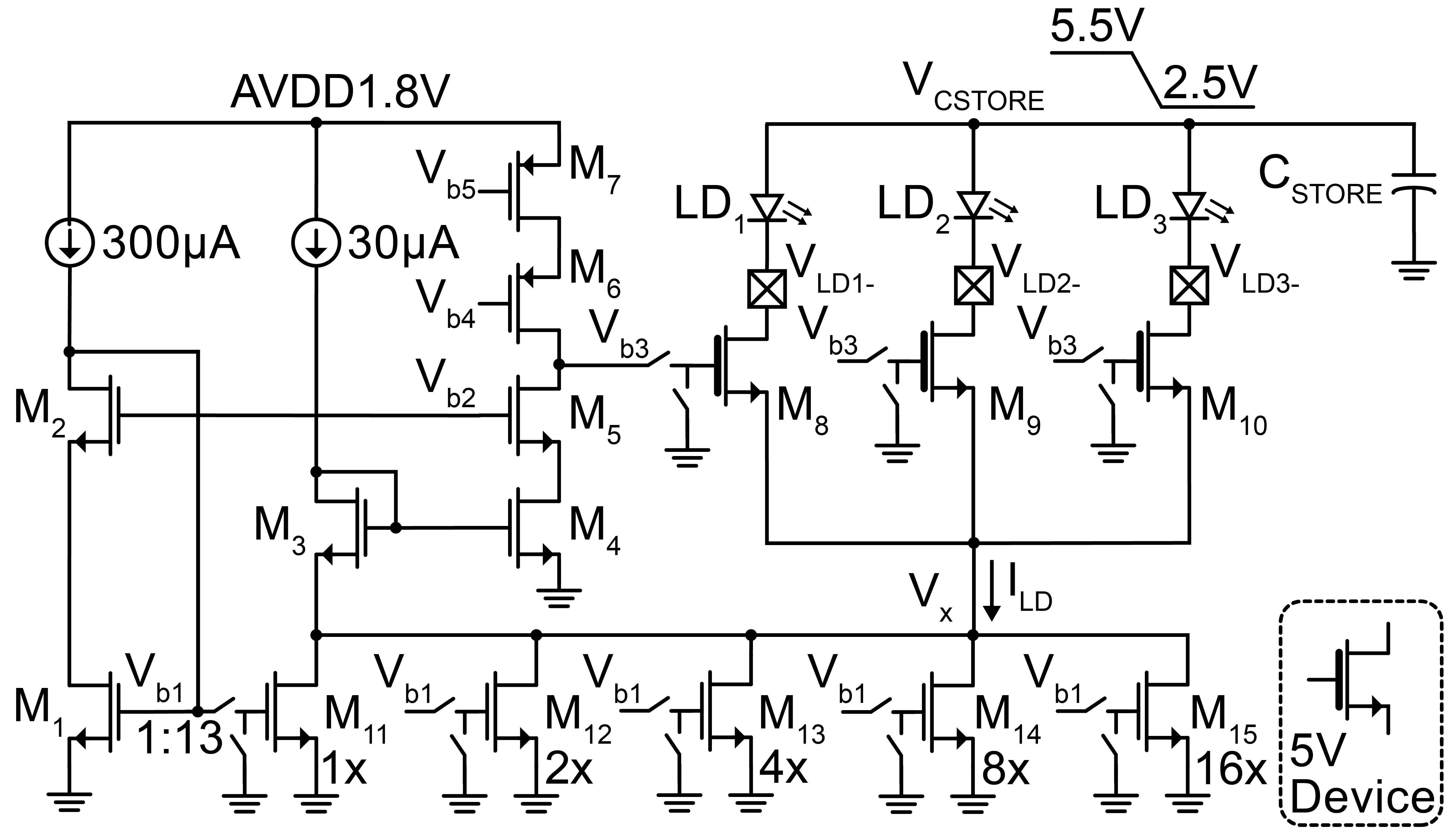}
\caption{Schematic of the 3-channel programmable laser driver.}
\label{laser driver}
\end{figure}

Fig. \ref{laser driver} shows the schematic of the 3-channel laser driver with programmable output current. To minimize the change in driver current, I\textsubscript{LD}, across the large voltage drop on V\textsubscript{CSTORE} (5.5–2.5 V), the driver must have high output impedance. Therefore, a gain-boosted cascode current source topology is used, in which the output impedance of the current source (M\textsubscript{8}–M\textsubscript{15}) is multiplied by the 65 dB gain of the cascoded boost amplifier (M\textsubscript{4}–M\textsubscript{7}). A 5-bit current DAC (M\textsubscript{11}–M\textsubscript{15}) enables a programmable output current from 0–115 mA with a 3.9 mA LSB. While the {\textmu}LDs in this work operate under 40 mA (see Fig. \ref{laser char}), this range accommodates a variety of commercial {\textmu}LDs with threshold currents up to 100 mA for future applications. Since only one laser is turned on at a time, the same driver circuitry is used for all three lasers. Thus, the cascode transistors select between the laser channels. For maximum output swing, V\textsubscript{x} is set by a level-shifting diode, M\textsubscript{3}, to bias M\textsubscript{11}–M\textsubscript{15} at the edge of triode. A headroom of at least 400 mV is required at the drains of M\textsubscript{8}–M\textsubscript{10} (V\textsubscript{LD\textminus}) to ensure operation in saturation.
\subsection{Imaging Frontend and Readout}
\begin{figure}[!t]
\centering
\includegraphics[width=3.49in]{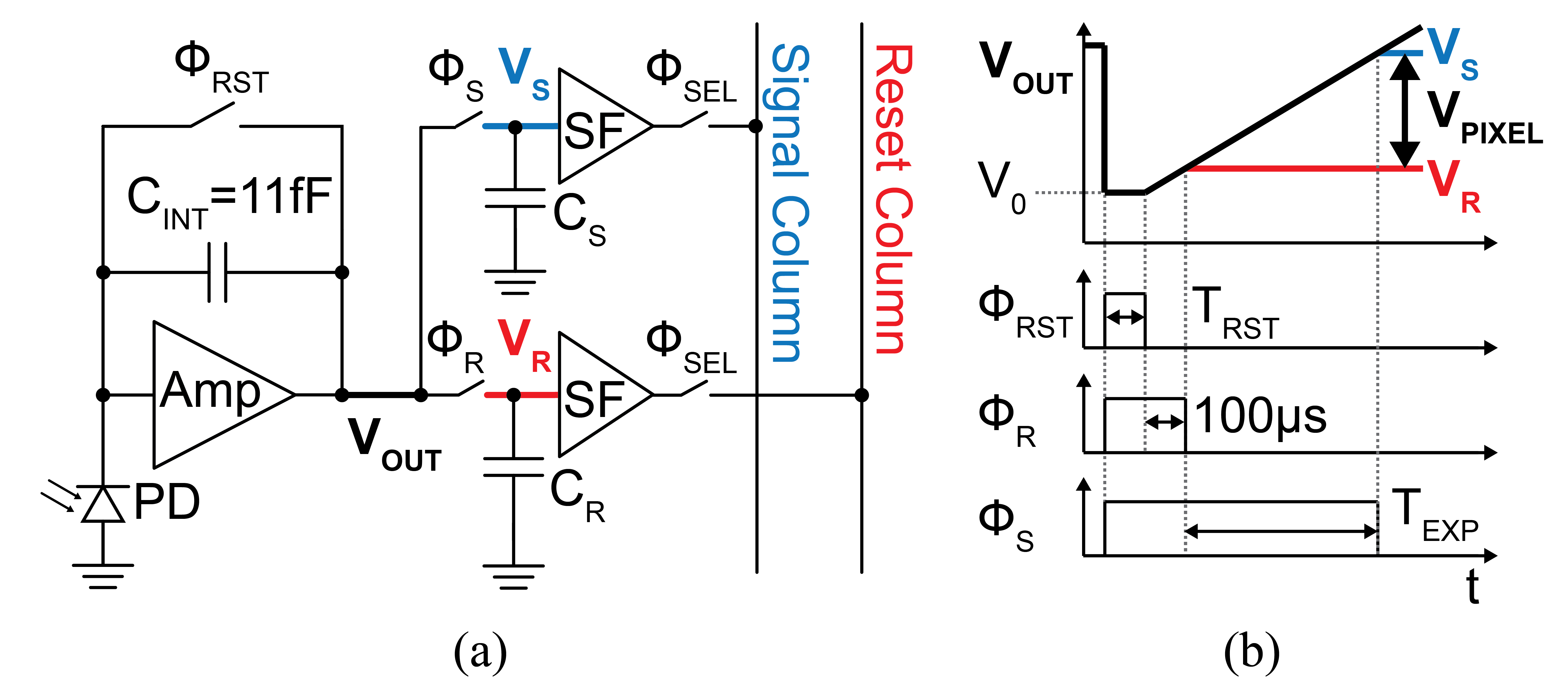}
\caption{(a) Active pixel architecture with correlated double sampling. (b) Pixel timing diagram.}
\label{pixel arch}
\end{figure}

The imaging frontend is similar to that presented in \cite{papageorgiou_chip-scale_2020}, but without the angle selective gratings as image deblurring is now provided by the FOP. The image sensor consists of a 36×40 array of pixels with a 44×44 {\textmu}m\textsuperscript{2} Nwell/Psub photodiode and a 55 {\textmu}m pitch, covering a 2×2.2 mm\textsuperscript{2} FoV.  The pixel architecture, shown in Fig. \ref{pixel arch}(a), is based on a CTIA with C\textsubscript{INT}=11 fF. To reduce low-frequency noise, reset switch sampling noise, and pixel offset, a correlated double-sampling scheme is implemented with the following pixel timing (illustrated in Fig. \ref{pixel arch}(b)). First, the voltage on C\textsubscript{INT} is set to zero during the initial reset phase, T\textsubscript{RST}, with timing configured in the \textit{Set T\textsubscript{EXP}} state. For the exposure time, T\textsubscript{EXP}, the photocurrent is integrated on C\textsubscript{INT} generating the pixel output voltage, $V_{OUT}=V_{0}+I_{PD}T_{EXP}/C_{INT}$, which is sampled on reset (C\textsubscript{R}) and signal (C\textsubscript{S}) sampling capacitors after intervals of 100 {\textmu}s and $T_{EXP}+100$ {\textmu}s, respectively. The final pixel value (V\textsubscript{PIXEL}) is the difference between the signal (V\textsubscript{S}) and reset (V\textsubscript{R}) values.

After \textit{Imaging}, the analog pixel values are digitized and stored in memory during the \textit{Readout} state. \textit{Readout} duration is set to limit the leakage on the in-pixel sampling capacitors to less than an LSB. Therefore, the readout is performed in parallel across 4 channels each spanning 10-pixel columns. Each channel consists of an 8-bit differential SAR ADC (Fig. S3) driven by a buffer. The ADC has a dynamic range of 500 mV with an LSB of 1.95 mV, which is below the pixel readout noise (see Section III.E). The readout circuits operate on a 3.3 V supply to ensure sufficient headroom considering that the in-pixel source followers level-shift the sampled pixel voltages up by 1 V.  Thus, the size of C\textsubscript{VCP} is chosen to maintain V\textsubscript{CP} above 3.5 V throughout this state. The signal (V\textsubscript{S}) and reset (V\textsubscript{R}) pixel values are subtracted by the differential ADCs, and the digitized pixel values are stored immediately after conversion in a 11.52 kb latched-based memory. Unlike the work in [32], this design enables a short \textit{Readout} time of 5.4 ms, which is not limited by the longer \textit{Backscattering} state (890 ms at 5 cm depth) that increases with depth due to the longer time of flight of the acoustic waves. 

\subsection{Data Transmission}
During \textit{Backscattering}, the memory is read serially ($\Theta$\textsubscript{MOD} in Fig. \ref{system diagram}) and transmitted by modulating the amplitude of the reflected (backscattered) US pulses using a switch (S\textsubscript{MOD} in Fig. \ref{system diagram}). The uplink communication protocol is shown in the timing diagram in Fig. \ref{timing diagram}. The transmitted data for each pixel comprises a 9-bit packet containing a header (set to 0) followed by 8 data bits. The header pulse allows for a one-pulse delay to make sure memory is read and loaded into the serializer before data transmission. Additionally, the header is set to a known value of zero to help identify the backscattered bit values. 

The external transducer generates a sequence of pulses each spanning a few cycles of the US carrier for the header and 8 individual bits. After a time of flight (ToF=33 {\textmu}s for 5 cm depth) the acoustic pulses reach the piezo and reflect with an amplitude proportional to the reflection coefficient of the piezo, $\Gamma$. $\Gamma$ is dependent on the electrical impedance loading the piezo, $R_{LOAD}$ and, therefore, can be controlled through the S\textsubscript{MOD} switch. Near the parallel resonance frequency of the piezo, $\Gamma \propto R_{PIEZO}/(R_{LOAD}+R_{PIEZO})$, where $R_{PIEZO}$ is the equivalent resistance of the piezo \cite{ghanbari_optimizing_2020}. The S\textsubscript{MOD}  switch impedance can be configured (hard-coded) by 2 bits to account for different $R_{PIEZO}$ values. After a second ToF, the backscattered signal is received by the external transducer and is demodulated to reconstruct the image. To avoid overlap of high voltage Tx and low voltage reflected Rx pulses, the external transducer transmits 2 bits within 2 ToFs and listens for the next 2 ToFs as shown in Fig. \ref{timing diagram}. 

\section{Measurement Results}
\begin{figure}[!t]
\centering
\includegraphics[width=3.49in]{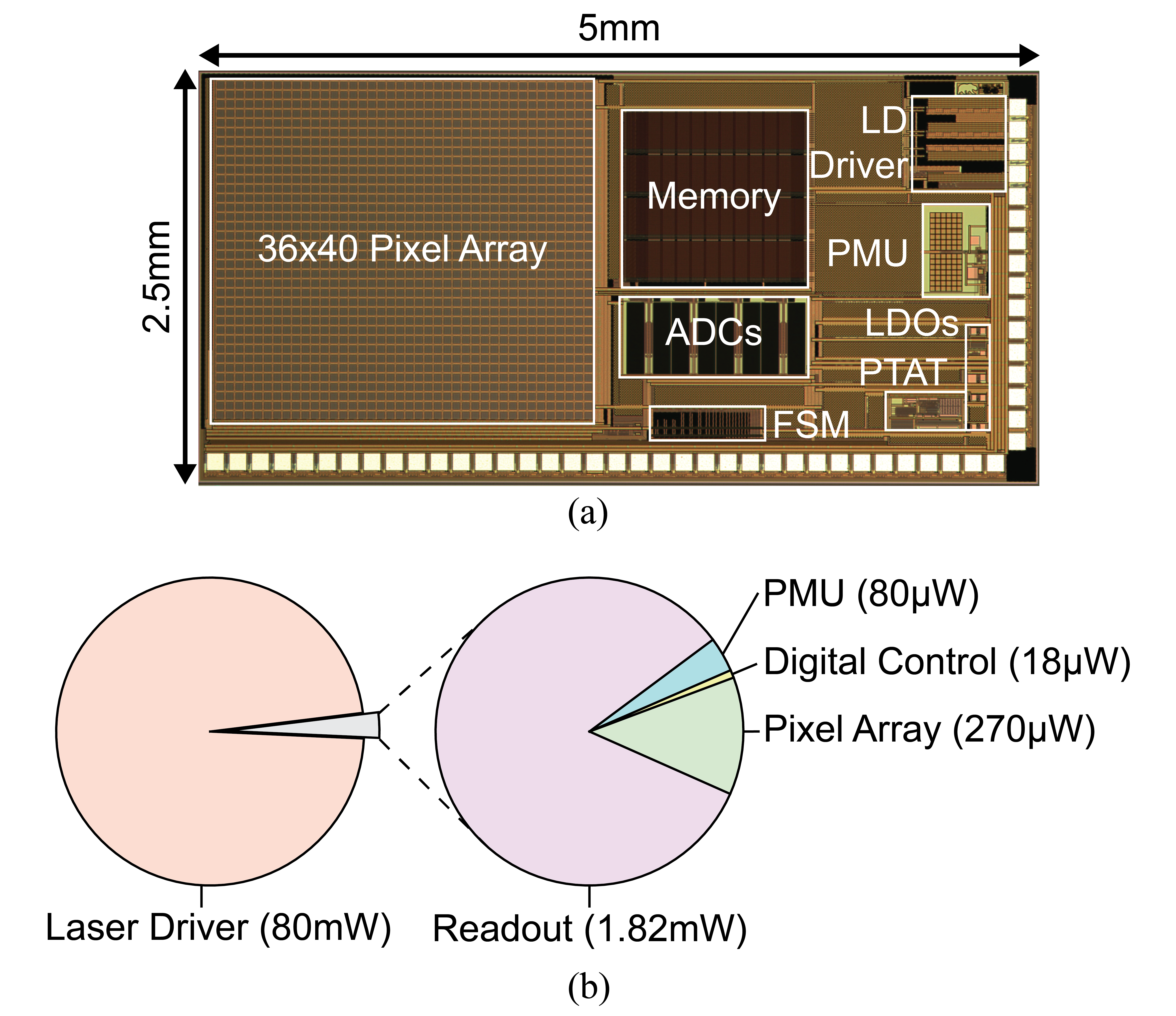}
\caption{(a) Chip micrograph. (b) Breakdown of system power consumption.}
\label{die photo}
\end{figure}

Fig. \ref{die photo}(a) shows the die photo of the chip. The ASIC measures 2.5×5 mm\textsuperscript{2} and is fabricated in a TSMC 180 nm high-voltage (1.8/5/32 V) LDMOS CMOS process. 1.8 V transistors are used for the digital, pixel, and laser driver, and 5 V devices are used for the PMU and pixel readout. Fig. \ref{die photo}(b) shows the power breakdown for the chip where the laser driver dominates the power consumption.

This section presents system-level measurement results for the US wireless link, laser driver, and imaging frontend.
\subsection{Measurement Setup}
\begin{figure}[!t]
\centering
\includegraphics[width=3.49in]{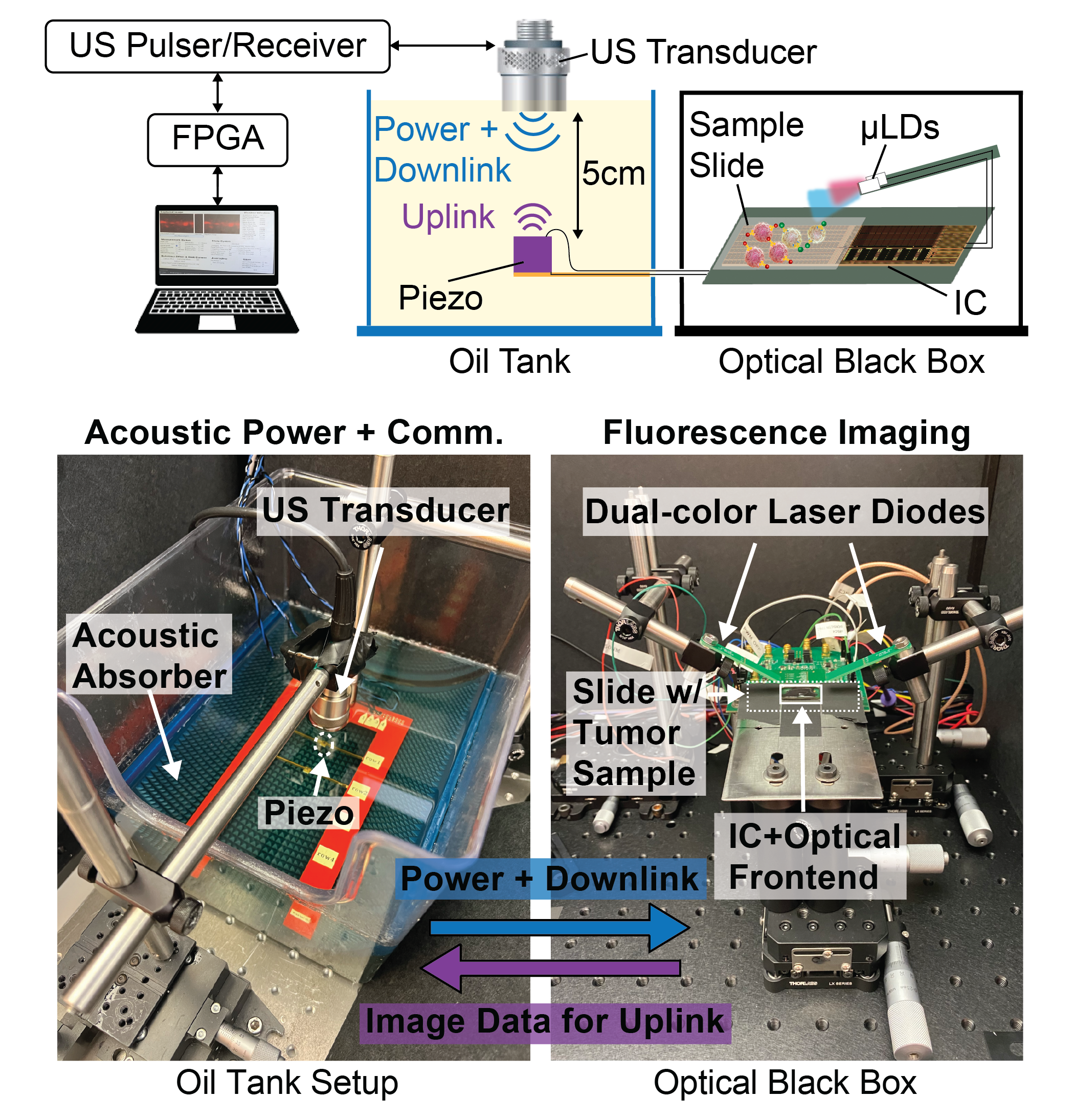}
\caption{Measurement setup for wireless imaging.}
\label{wireless imaging setup}
\end{figure}

Fig. \ref{wireless imaging setup} shows the measurement setup for demonstrating fully wireless operation of the chip. In the acoustic setup, the piezo is submerged at 5 cm depth in a tank of canola oil. An external focused transducer (V314-SU-F1.90IN-PTF, Evident Scientific) at the surface of the tank transmits US signals to the piezo. To minimize interference from US reflections on data uplink, an acoustic absorber (Aptflex F28P, Precision Acoustics) is placed at the bottom of the tank. An FPGA (Opal Kelly, XEM7010) generates the desired US pulse sequence (as in Fig. \ref{timing diagram}) to control the chip. The timing of the pulse sequence is programmed through a custom user interface that interfaces with the FPGA. The generated waveforms are sent to a high-voltage transducer pulser board (Max14808, Maxim Integrated) to drive the external transducer accordingly. 

The chip is directly connected with wires to the piezo for wireless power harvesting and data transfer via US. It is located inside a black box to reduce the background signal from ambient light during imaging. Slide-mounted samples are placed directly on top of the chip. The chip drives the {\textmu}LDs, mounted on separate PCBs, to transilluminate the sample from above. Admittedly, \textit{in vivo}, the sample must be epi-illuminated between the sensor and the tissue. Epi-illumination can be accomplished in the future by directing the laser light through a glass separator or light guide plate placed on top of the sensor \cite{sasagawa_front-light_2023, shin_miniaturized_2023}.

After taking an image, the backscattered US pulses are received by the external transducer and captured on an oscilloscope for processing and demodulation. To remove the pixel-to-pixel DC offsets due to the photodiode dark current and mismatch in the readout circuitry, a dark image with the same integration time but with the laser off is subtracted from the final fluorescence image. The dark image is averaged to minimize its noise contribution.

\subsection{Ultrasound Wireless Power Transfer}
\begin{figure*}[!t]
\centering
\includegraphics[width=7.14in]{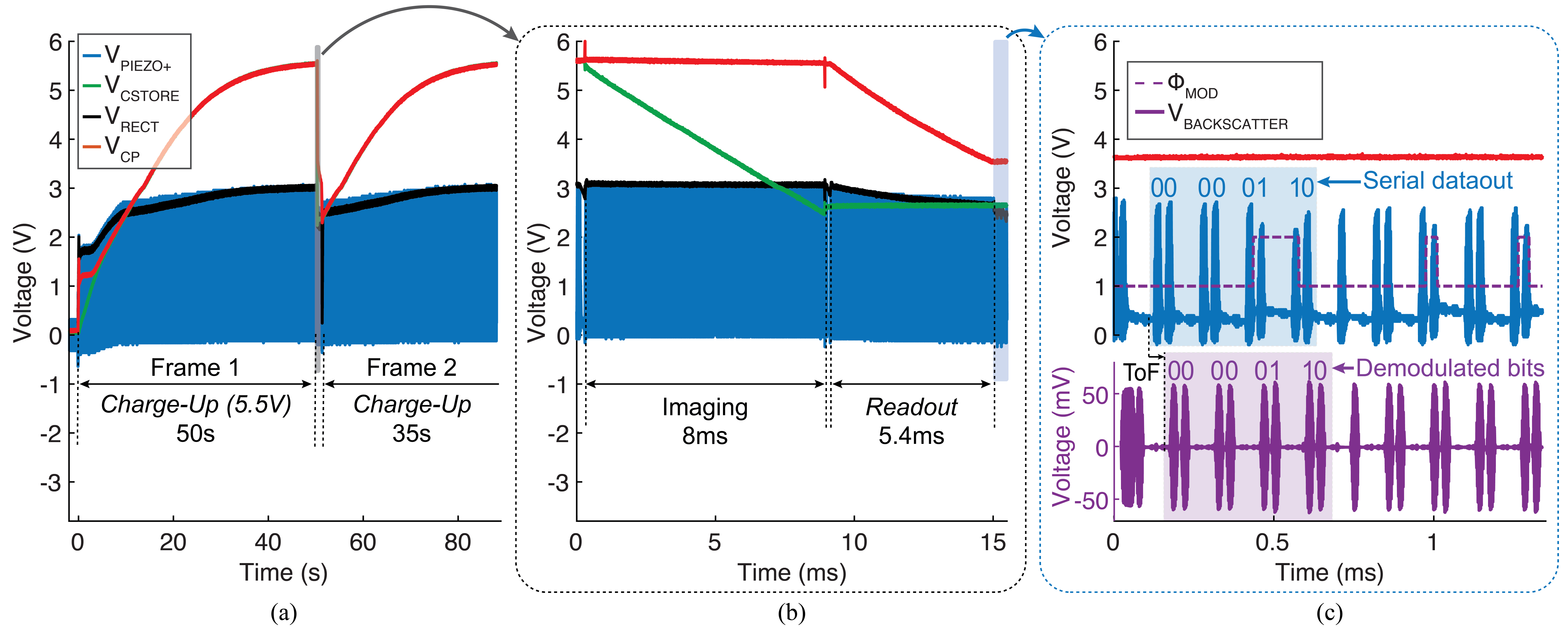}
\caption{Measured PMU waveforms during (a) \textit{Charge-Up} and (b) \textit{Imaging} and \textit{Readout}. (c) Measured backscatter waveforms.}
\label{measured timing}
\end{figure*}

Fig. \ref{measured timing}(a) shows the measured PMU waveforms (V\textsubscript{PIEZO+}, V\textsubscript{RECT}, V\textsubscript{CP}, V\textsubscript{CSTORE}), verifying wireless operation of the full system at 5 cm depth. In this measurement, the system operates with an US power density of 221 mW/cm\textsuperscript{2}, which falls within 31\% of FDA safety limits. Under this minimum required acoustic power condition, V\textsubscript{CP} charges to 5.5 V in 50 s for the initial image. The charging time decreases to 35 s for consecutive frames with a nonzero initial V\textsubscript{CP}. The \textit{Charge-Up} time can be further reduced by increasing US power intensity, operating closer to the FDA limits. The output voltages of the rectifier (V\textsubscript{RECT}) and charge pump ((V\textsubscript{CP})) across different input voltages (V\textsubscript{PIEZO+}) show a minimum V\textsubscript{PIEZO+}=2.42 V is required for stable operation of the chip (Fig. S4).

Measured PMU waveforms during the \textit{Imaging} and \textit{Readout} states are presented in Fig. \ref{measured timing}(b). During \textit{Imaging} (T\textsubscript{EXP}=8 ms), V\textsubscript{CSTORE} drops from 5.5–2.5 V while supplying the laser with I\textsubscript{LD}=37.5 mA from the energy stored in C\textsubscript{STORE}. V\textsubscript{CP} remains at 5.5 V throughout \textit{Imaging} and drops to 3.5 V during \textit{Readout}. 

Fig. \ref{timing diagram}(c) shows the measured waveforms while transmitting a single pixel data packet via US backscattering. V\textsubscript{PIEZO+} is modulated according to the serial output of the memory ($\Theta$\textsubscript{MOD}) and the backscattered pulses are received by the external transducer (V\textsubscript{BACKSCATTER} in Fig. \ref{timing diagram}(c)). The one bits correspond to a smaller load impedance, but appear larger in amplitude than the zero bits because the piezo is operated between series and parallel resonance frequencies for maximum voltage harvesting. 

\begin{figure}[!t]
\centering
\includegraphics[width=3.49in]{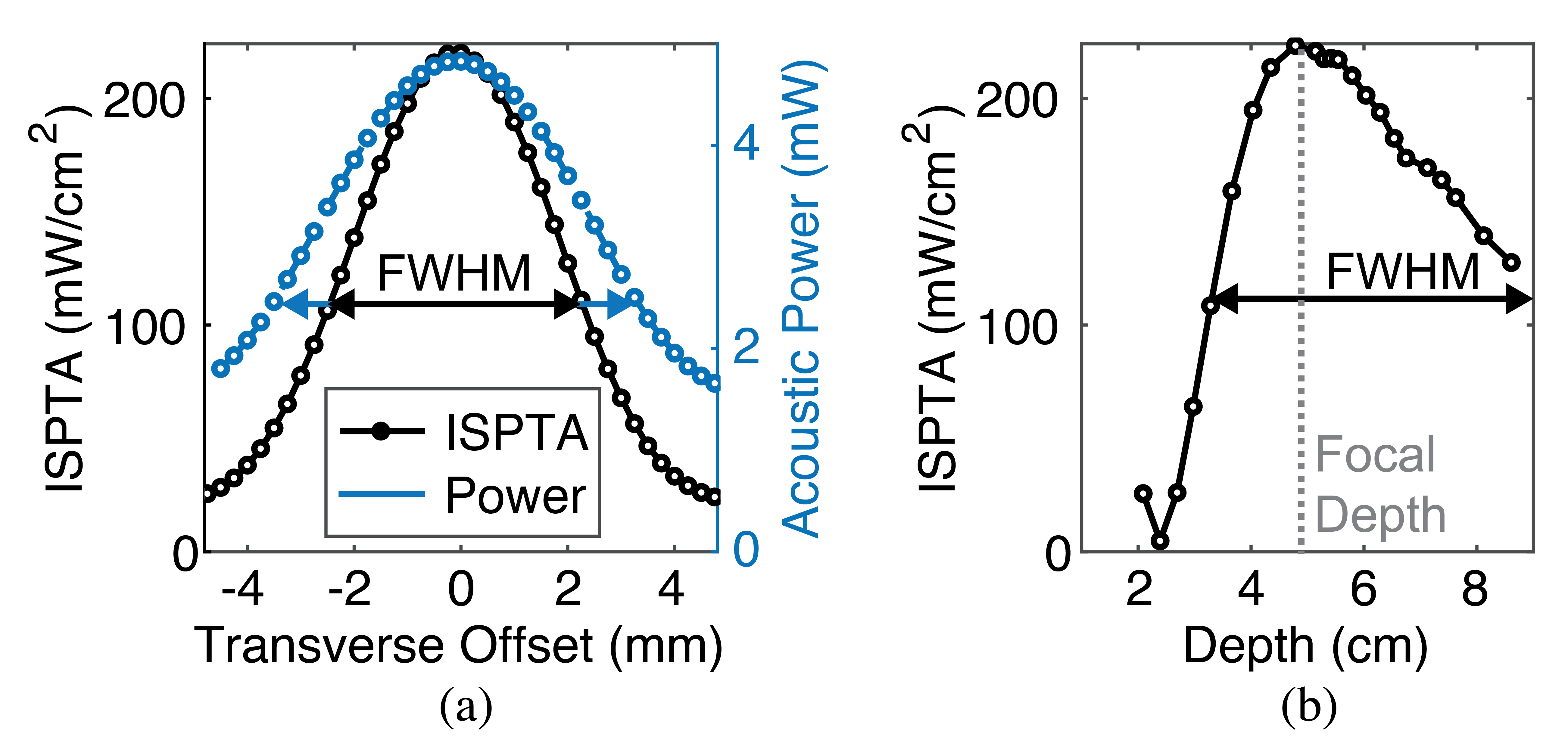}
\caption{Harvested acoustic power vs. (a) transverse offset and (b) depth.}
\label{US alignment}
\end{figure}

Fig. \ref{US alignment}(a) shows the total acoustic power and acoustic power density (ISPTA) incident on the piezo surface area at 5 cm depth for transverse offsets along the X or Y axis. Fig. \ref{US alignment}(b) shows a similar measurement as the depth is adjusted along the Z axis. The acoustic power density is measured with a hydrophone (HGL-1000, Onda) and it is integrated over the piezo area to measure the available acoustic power at the piezo surface. The reported spatial-peak time-average intensity (ISPTA) of the acoustic field is the relevant parameter in calculating FDA safety limits for diagnostic US \cite{health_marketing_2023}. For both transverse and depth offsets, the power decreases as the piezo moves away from the focal point (near 5 cm depth) of the external transducer. The measured transverse and axial FWHMs for ISPTA are 4.5 mm and 60 mm, respectively. In the future, misalignment loss can be reduced through dynamic focusing of the US with beam forming \cite{benedict_phased_2022}. It should be noted that angular misalignment of the piezo with respect to the US beam will also reduce the harvested power \cite{sonmezoglu_monitoring_2021, piech_wireless_2020}.

While charging V\textsubscript{CP} from 0–5.5 V, the overall electrical energy efficiency of the PMU is 12.7\%. The efficiency of the system in converting the available acoustic energy on the face of the piezo to the electrical output energy of the PMU is 3.3\%. The output energy of the PMU is calculated by measuring the energy stored in the C\textsubscript{STORE} and C\textsubscript{VCP} and the total energy consumption of the ASIC during Charge-Up. The input acoustic energy is calculated by integrating the measured acoustic power density at the surface of the piezo (Fig. \ref{US alignment}(a)) throughout this same period. 

\subsection{Ultrasound Data Uplink}
At 5 cm depth, transmission of one image (11.52 kb) takes 890 ms, resulting in a data rate of 13 kbps. The received backscattered waveform is processed and demodulated to reconstruct the image as follows. First, the signal is bandpass-filtered at the carrier frequency, windowed to select the bit intervals, and then reconstructed with sinc interpolation. The peak-to-peak amplitude is then measured for each pulse and compared with a predetermined threshold to predict the bit value. The serial output of the chip serves as the ground truth. 

\begin{figure}[!t]
\centering
\includegraphics[width=3.49in]{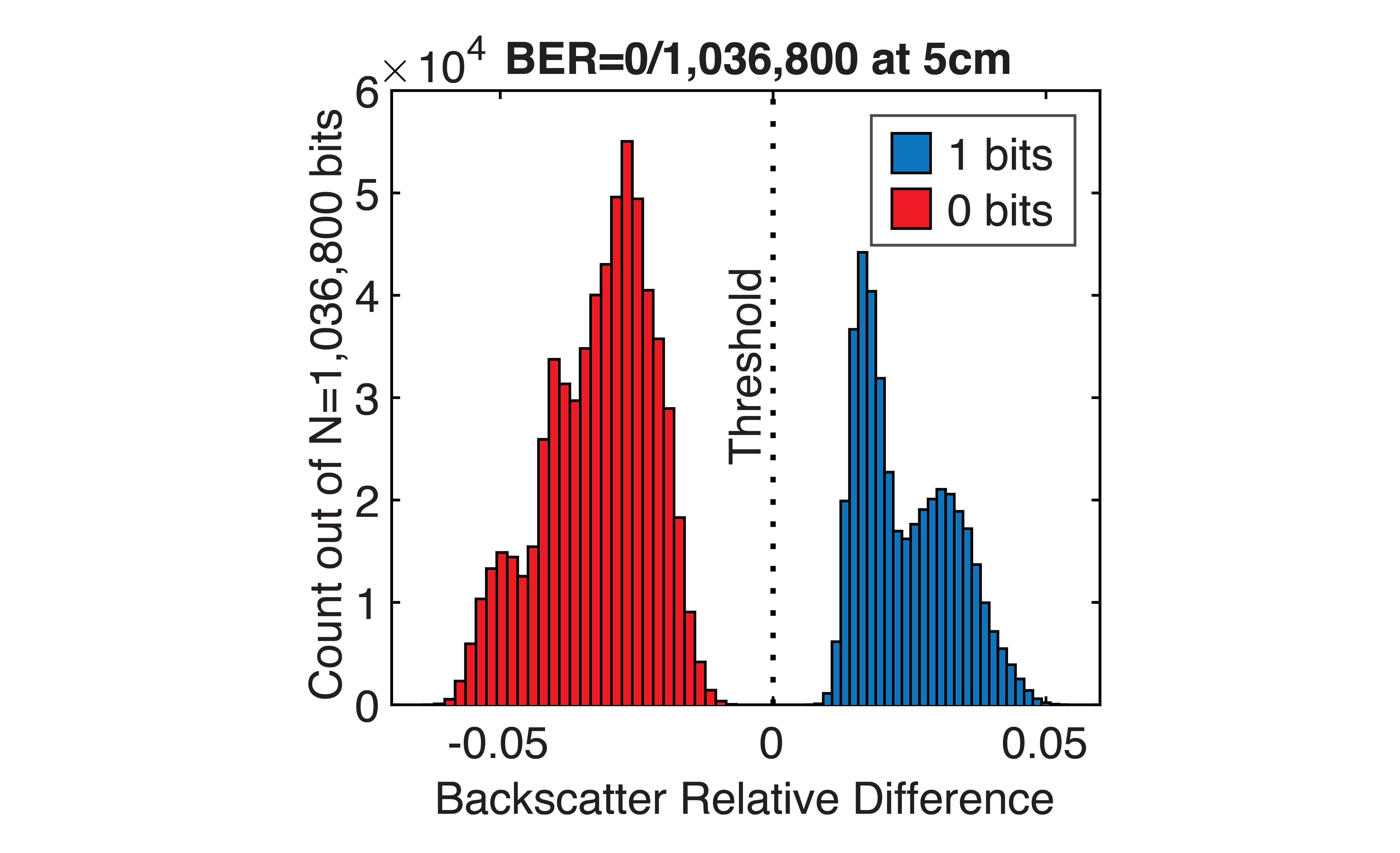}
\caption{Measured bit error rate (BER) at 5cm depth in oil.}
\label{BER}
\end{figure}

Fig. \ref{BER} shows a histogram of the backscattered signal amplitude for each bit normalized to the threshold amplitude, demonstrating a clear separation between one and zero bits. The measurement shows robust error-free transmission of 90 frames, including a combination of dark frames and images taken with the 650 nm and 455 nm lasers. The bimodal nature of the histogram results from combining data across different imaging conditions and differing interference from the high voltage pulsing of the external transducer on the two pulses received within each interval of 2 ToFs.  The device achieves a BER better than 10\textsuperscript{-6} (0 out of 1,036,800 bits) with an average modulation index of 5.6\%.

\subsection{Laser Driver}
\begin{figure}[!t]
\centering
\includegraphics[width=3.49in]{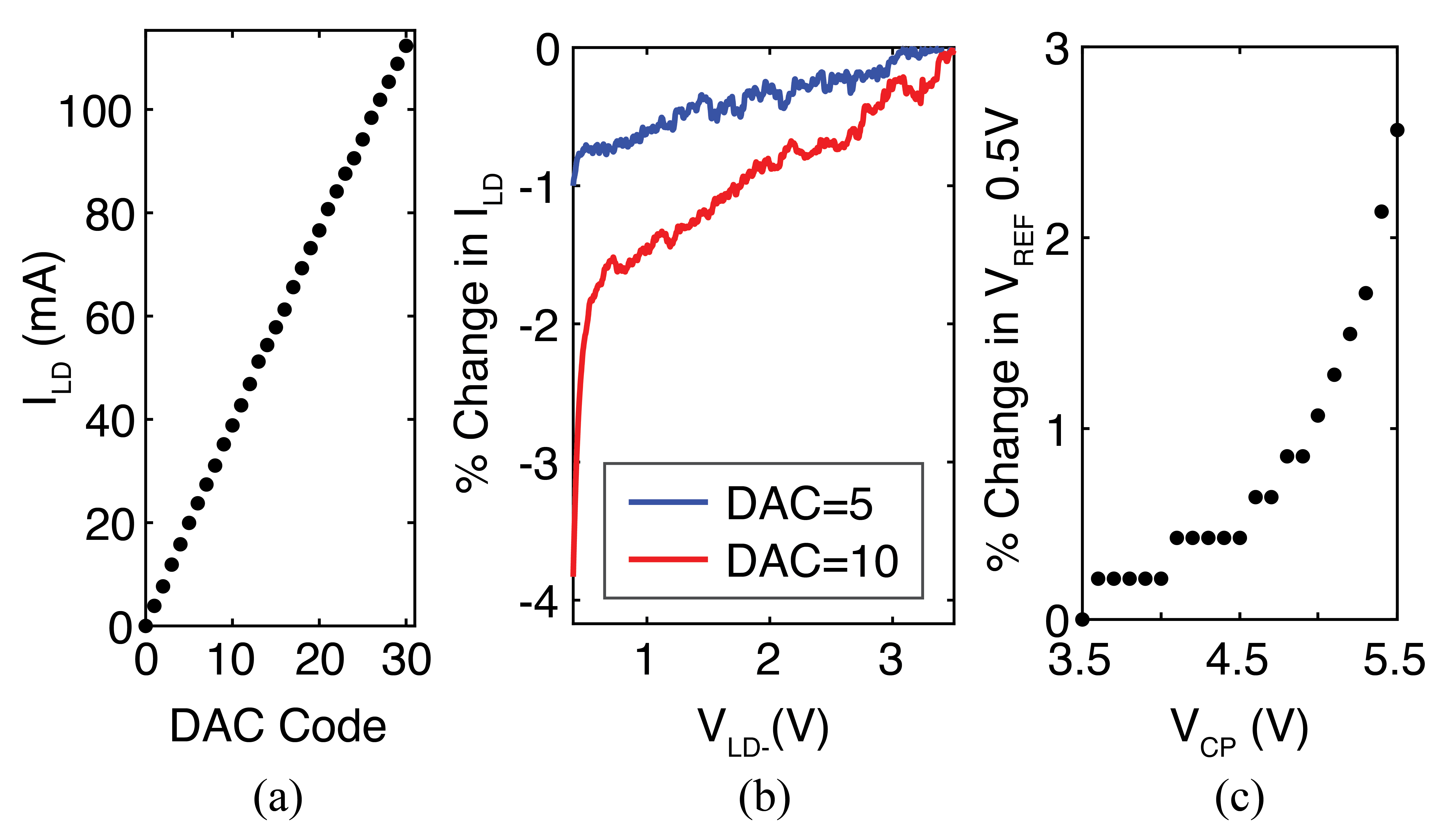}
\caption{Measurements of (a) laser driver current (I\textsubscript{LD}) vs. DAC code, (b) percent change in I\textsubscript{LD} vs. driver output voltage (V\textsubscript{LD-}), and (c) PTAT voltage reference (V\textsubscript{REF} 0.5 V) variation vs. V\textsubscript{CP}.}
\label{laser driver measurements}
\end{figure}

Fig. \ref{laser driver measurements} shows measurements of the laser driver and PTAT. The output current of the laser driver (I\textsubscript{LD}) is measured with a precision measurement unit (B2912A, Keysight). Fig. \ref{laser driver measurements}(a) shows the measured I\textsubscript{LD} across all DAC codes and Fig. \ref{laser driver measurements}(b) shows the percent change in I\textsubscript{LD} as the output voltage of the laser driver, V\textsubscript{LD\textminus}, drops from 3.5–0.4 V. This range corresponds to the V\textsubscript{LD\textminus} for a 5.5–3.5 V drop on V\textsubscript{CP} accounting for the 2 V forward bias voltage of the 650 nm {\textmu}LD. For DAC=5 (I\textsubscript{LD}=20 mA), there is less than 1\% variation across the 3.1 V drop, corresponding to 1.3\% variation in optical power output of the 650 nm {\textmu}LD.

Fig. \ref{laser driver measurements}(c) shows the variation in the 0.5 V PTAT reference across V\textsubscript{CP} measured through the V\textsubscript{ADC}0.5V LDO. As V\textsubscript{CP} drops from 5.5–3.5 V, the PTAT reference varies around 2.5\%, which has minimal effect on the ADC during \textit{Readout}.

These results are an improvement over \cite{rabbani_towards_2024} where the reference current varied 11.5\% over a 1.5 V drop, resulting in a 50\% reduction in the laser output power.

\subsection{Imaging Frontend}
\begin{figure}[!t]
\centering
\includegraphics[width=3.49in]{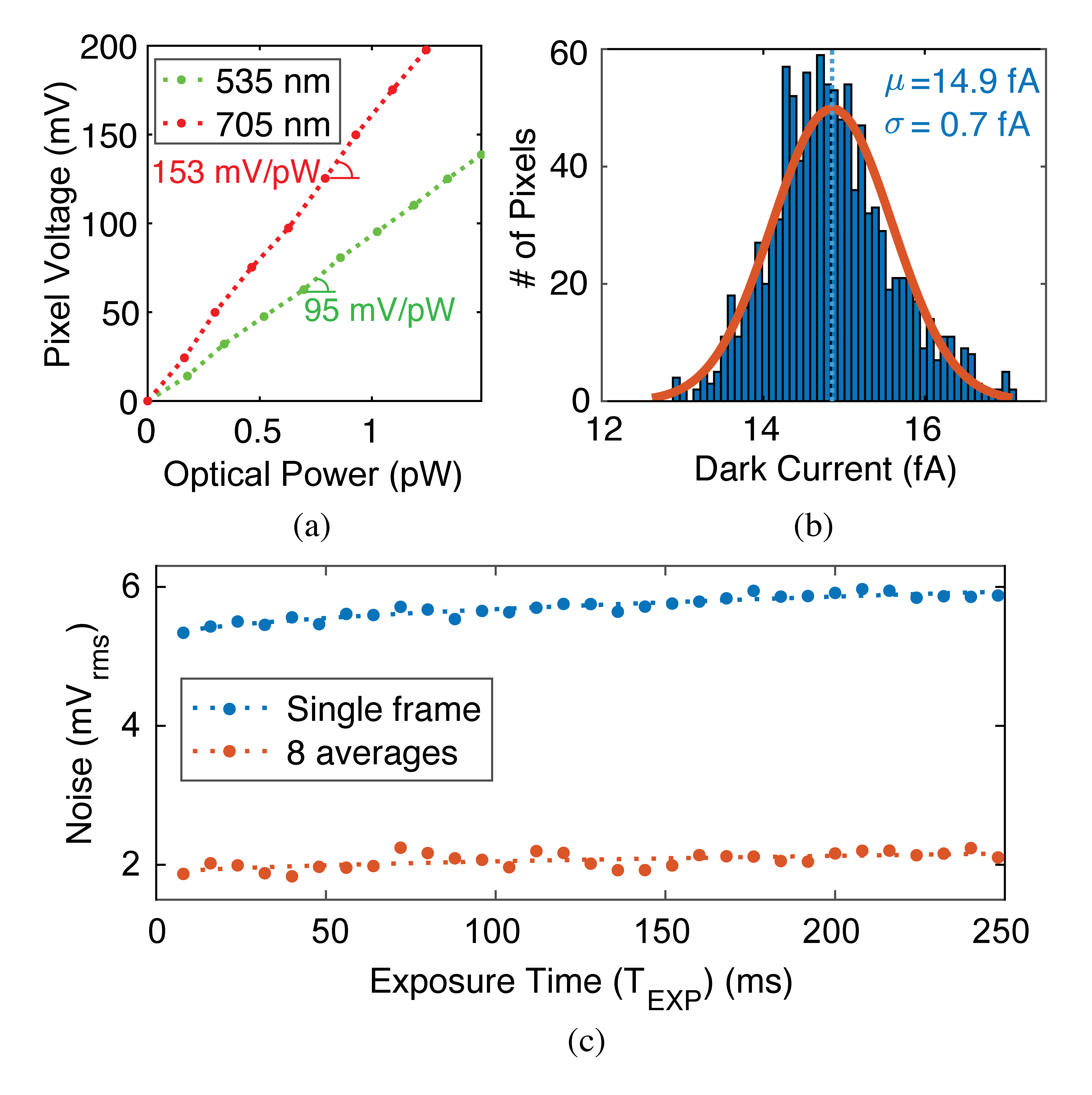}
\caption{(a) Pixel output voltage vs. incident optical power. (b) Histogram of measured dark current across pixels. (c) Measured pixel noise under dark condition for a single frame and after 8 averages.}
\label{pixel char}
\end{figure}

The photodiode responsivity is determined by measuring pixel output voltage across a range of incident optical powers as shown in Fig. \ref{pixel char}(a). We use a LED with a collimator and beam expander to ensure uniform illumination of the sensor. A narrow bandpass interference filter placed in front of the LED selects a specific wavelength. Measurements are made at 535 nm and 705 nm, near the center of the optical frontend passbands. The optical power output of the LED is characterized with a power meter (PM100D, ThorLabs). In Fig. \ref{pixel char}(a), the slope indicates pixel gain in mV/pW with T\textsubscript{EXP}=8 ms. The photodiode responsivity is calculated by dividing pixel gain by the transimpedance gain of the CTIA. The pixels have a mean responsivity of 0.13 A/W (quantum efficiency (QE=30\%) and 0.21 A/W (QE=37\%), at 535 nm and 705 nm respectively.

A histogram of the measured dark current across pixels with a Gaussian fit is shown in Fig. \ref{pixel char}(b). The mean dark current is 14.9 fA (7.7 aA/{\textmu}m\textsuperscript{2}) with a standard deviation of 0.7 fA (0.4 aA/{\textmu}m\textsuperscript{2}). Fig. \ref{pixel char}(c) shows the measured pixel output noise in dark condition for different exposure times for a single frame and an average of 8 frames. For T\textsubscript{EXP}=8 ms, the measured pixel output noise is 5.34 mV\textsubscript{rms} for a single frame and  1.87 mV\textsubscript{rms} after 8 averages. The output noise increases with the exposure time due to the shot noise from the increased dark signal.

\begin{figure}[!t]
\centering
\includegraphics[width=3.49in]{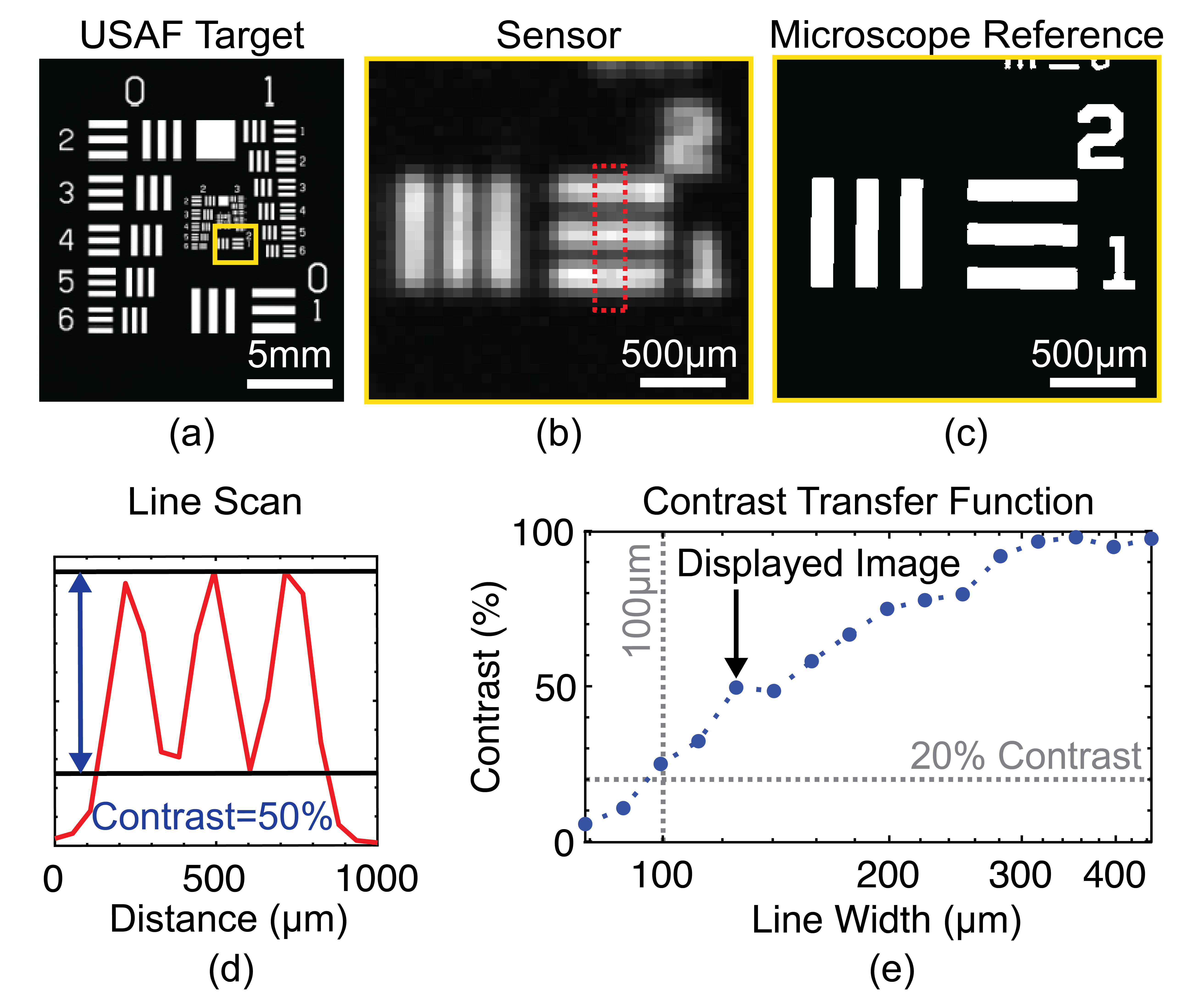}
\caption{Resolution measurements using (a) USAF target. Image of element with 125 {\textmu}m line width with the sensor (b) and a microscope (c). (d) Line scan of image in (a). (e) Measured contrast transfer function.}
\label{resolution}
\end{figure}

The resolution of the imager is measured with a negative USAF target (Fig. \ref{resolution}(a)) overlaying a uniform layer of Cy5 NHS ester ({$\lambda$}\textsubscript{EX}=649 nm, {$\lambda$}\textsubscript{EM}=670 nm) dissolved in PBS at 10 {\textmu}M concentration. The dye is contained with a 150 {\textmu}m-thick glass coverslip and the target is placed on the imager. The resolution measurements were conducted with wired power and data transfer and using a fiber-coupled 650 nm laser for uniform illumination. Fig. \ref{resolution}(b) shows the sensor image of the element with 125 {\textmu}m line spacing compared to the microscope reference image in Fig. \ref{resolution}(c). The sensor images this element at 50\% contrast as calculated with the line scan in Fig. \ref{resolution}(d). Contrast is calculated as $(V_{MAX}-V_{MIN})/(V_{MAX}+V_{MIN}-V_{BK})$, where $V_{MAX}$ and $V_{MIN}$ are the maximum and minimum pixel values in the bright and dark bars, respectively, and $V_{BK}$ is the background signal. Fig. \ref{resolution}(e) shows the full contrast transfer function measured by imaging elements on the target with line spacing ranging from 79–455 {\textmu}m and calculating the contrast for each. These results demonstrate that with the FOP, the imager can distinguish line spacing as small as 100 {\textmu}m with greater than 20\% contrast. 

\begin{figure}[!t]
\centering
\includegraphics[width=3.49in]{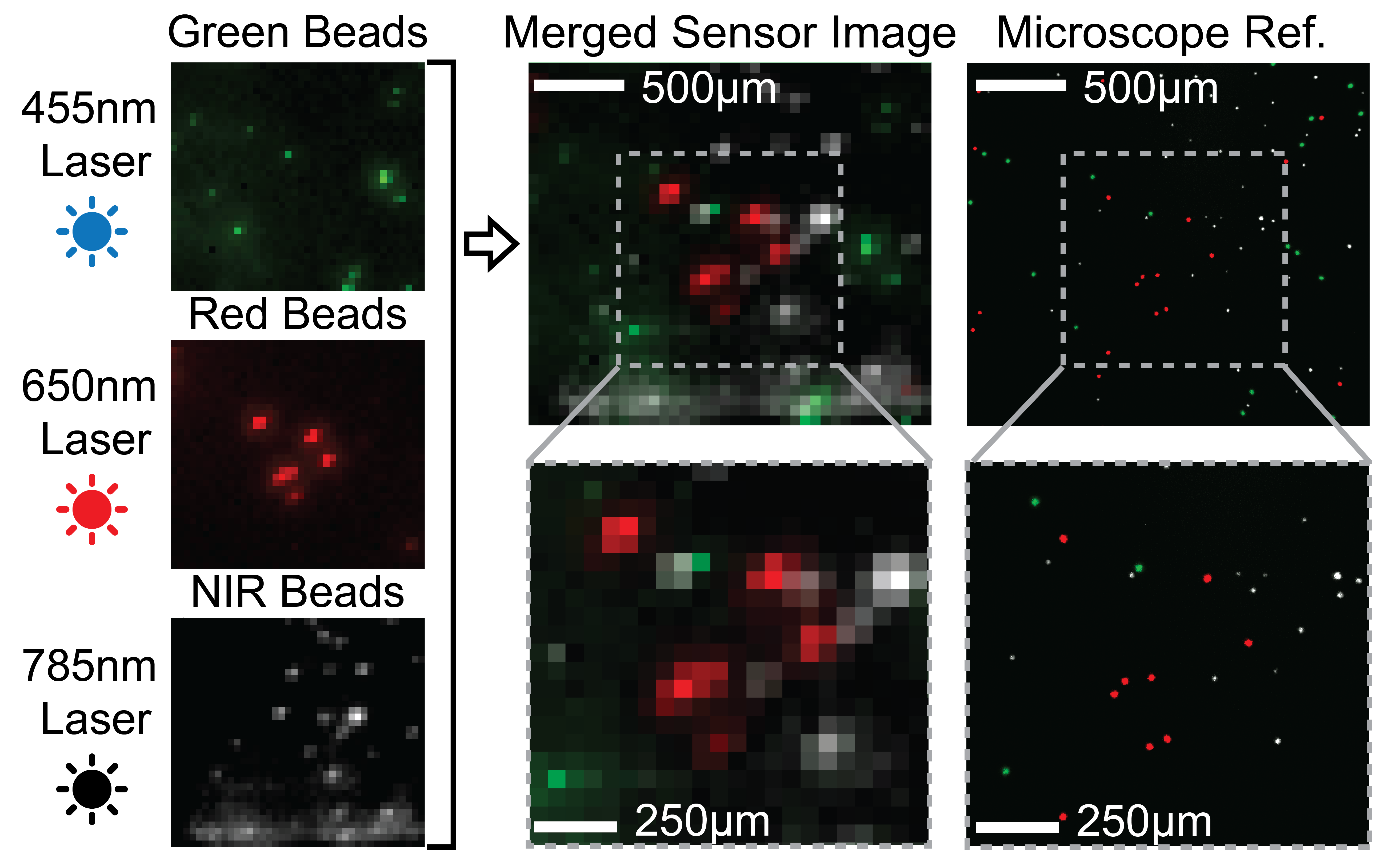}
\caption{3-color imaging of fluorescent beads.}
\label{three-color imaging}
\end{figure}

To demonstrate three-color imaging, we image a sample containing 15{\textmu}m-diameter green ({$\lambda$}\textsubscript{EX}=505 nm, {$\lambda$}\textsubscript{EM}=515 nm, F8844, Thermo Fisher Scientific), red ({$\lambda$}\textsubscript{EX}=645 nm, {$\lambda$}\textsubscript{EM}=680 nm,  F8843, Thermo Fisher Scientific), and NIR ({$\lambda$}\textsubscript{EX}=780 nm, {$\lambda$}\textsubscript{EM}=820 nm, DNQ-L069, CD Bioparticles) fluorescent beads. The beads are suspended in 1× PBS solution at a concentration of approximately 10 beads/{\textmu}L. 50 {\textmu}L of solution is pipetted into a micro-well chamber slide for imaging. Imaging results are shown in Fig. \ref{three-color imaging}. The sensor images are obtained wirelessly with I\textsubscript{LD}=18.5 mA, T\textsubscript{EXP,GREEN}=8 ms, T\textsubscript{EXP,RED}=16 ms, T\textsubscript{EXP,NIR}=8 ms. For each color channel, 4 frames are averaged and the channels are colored and overlaid to make the multicolor image. The sensor images show good correspondence with the reference image taken with a bench-top fluorescence microscope (Leica DM-IRB). A few beads do not appear in the sensor image due to non-uniform illumination from the {\textmu}LDs. There is also a line artifact visible in the NIR channel due to reflections off the wire-bonds and that be mitigated through more careful fabrication as detailed in \cite{roschelle_multicolor_2024}.

\section{\textit{Ex Vivo} Imaging of Immune Response}
We conducted an \textit{ex vivo} mouse experiment to demonstrate the application of our sensor to assessing the response to cancer immunotherapy through dual-color fluorescence imaging of both effector and suppressor cells in the tumor microenvironment. In this study, we measure response to immune checkpoint inhibitors (ICIs), a class of immunotherapy that activates the immune system against cancer by blocking interactions between effector and inhibitory immune cells and cancer \cite{ribas_cancer_2018, robert_decade_2020}. A successful immune response to ICIs requires the activation and proliferation of CD8+ T-cells, the most powerful effector cells in the anticancer response, into the tumor microenvironment \cite{raskov_cytotoxic_2021}. Therefore, CD8+ T-cell infiltration has been identified as an indicator of a favorable immune response \cite{spitzer_systemic_2017}. However, CD8+ T-cell activation can be inhibited by suppressor immune cells such as neutrophils, which regulate the immune system and inflammation in the body and are associated with resistance to ICI immunotherapy \cite{faget_neutrophils_2021, kargl_neutrophil_2019}. Dual-color fluorescence imaging enables a differential measurement of these two control mechanisms of the immune response with the same imaging frontend which is not possible with clinical imaging modalities such as MRI, PET, or CT.

\subsection{Experimental Design}
Fig. S5 outlines the \textit{ex vivo} experiment design, which uses two engineered cancer models from \cite{woo_cho_t_2022}, an LLC lung cancer model (engineered to resist ICIs) and a B16F10 melanoma model (engineered to respond to ICIs). Both tumor models show increased CD8+ T-cell infiltration over the course of treatment. However, while the B16F10 tumors reliably respond, the LCC tumors are resistant to ICI therapy. This resistance has been linked to a T-cell-driven inflammatory response that triggers an influx of neutrophils into the tumor, suppressing T-cell activation \cite{woo_cho_t_2022}. 

The experiment includes two groups of mice each bearing one type of tumor. Each group consists of a mouse treated with a combination of PD-1 and CTLA-4 inhibitors, a class of ICIs  \cite{ribas_cancer_2018}, and an untreated mouse injected with a non-therapeutic antibody for control. Three weeks after tumor implantation, the tumors are harvested, sectioned to 4 {\textmu}m-thick samples, and mounted on glass slides. Two adjacent sections from each tumor are labeled separately with fluorescent probes targeting CD8+ T-cells and neutrophils. CD8+ T-cells are stained with a CD8+ antibody labeled with Cy5 ({$\lambda$}\textsubscript{EX}=649 nm, {$\lambda$}\textsubscript{EM}=670 nm) and neutrophils are stained with a CD11b antibody labeled with FAM ({$\lambda$}\textsubscript{EX}=492 nm, {$\lambda$}\textsubscript{EM}=518 nm). 
\begin{figure*}[!t]
\centering
\includegraphics[width=7.14in]{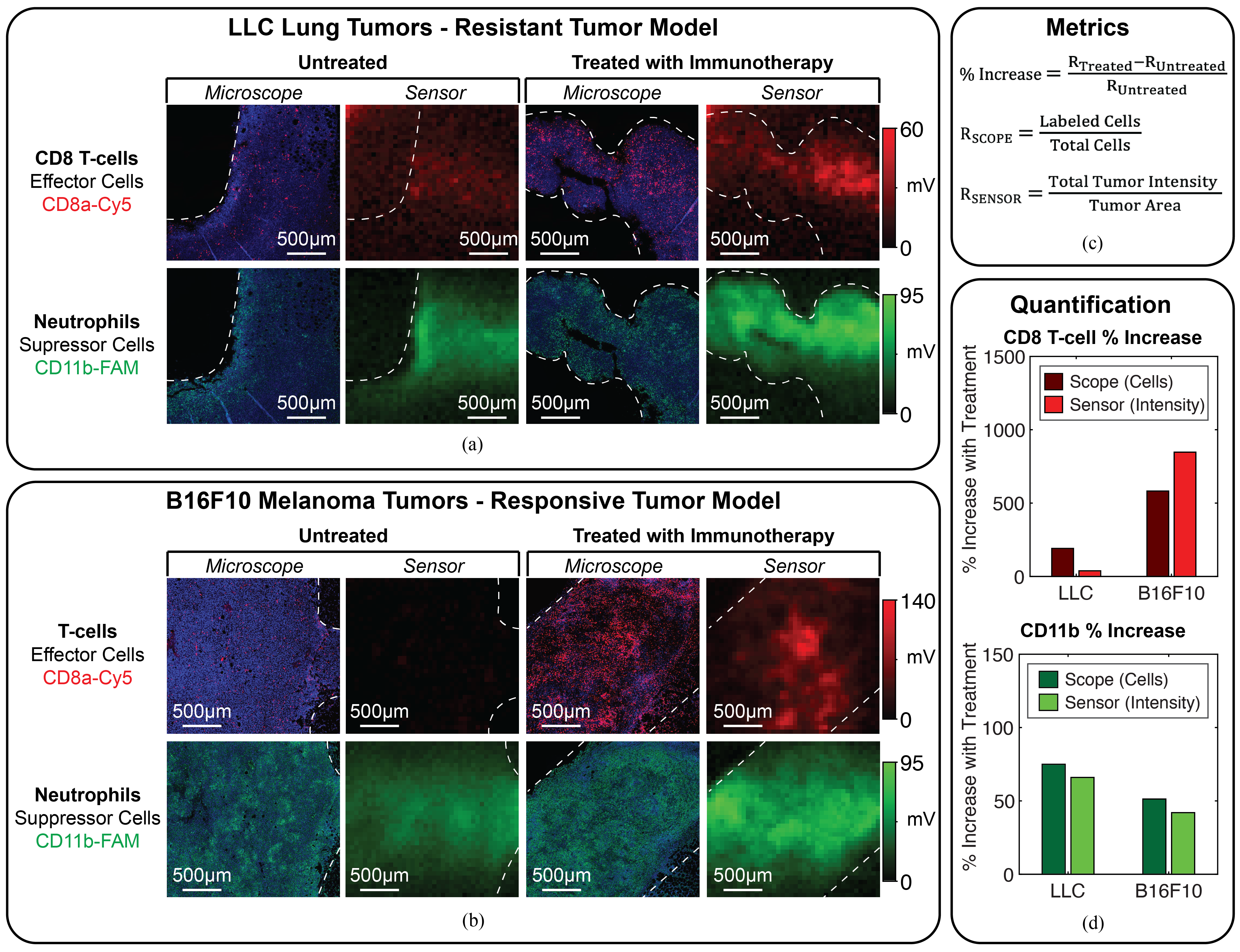}
\caption{\textit{Ex vivo} imaging of mouse tumors with and without immunotherapy. Imaging results for (a) the resistant tumor model (LLC) and (b) the responsive model (B16F10). (c) Metrics for quantification of cell populations. (d) Quantified results. }
\label{ex vivo imaging}
\end{figure*}

\subsection{Imaging Results}
Images of the tumor samples are captured wirelessly with the sensor and compared with reference images from a bench-top fluorescence microscope. Figs. \ref{ex vivo imaging}(a) and (b) show the imaging results from the LLC (resistant) and B16F10 (responsive) groups, respectively. For each fluorescent channel, 8 frames are acquired with the chip, using imaging parameters of I\textsubscript{LD}=18.5 mA, T\textsubscript{EXP,Cy5}=16 ms, and T\textsubscript{EXP,FAM}=8 ms. The sensor images are averaged across all frames. The microscope images are overlaid with the cell nuclei of the entire sample, stained with DAPI (blue in the image) to highlight the tumor area. The white lines within the images indicate the boundaries of the tumor tissue. The sensor images are qualitatively consistent with the microscope references, albeit at a lower resolution and with varying intensity across the image due to non-uniform illumination from the {\textmu}LDs.

To quantify the results for each tumor model, the percent change in the density of both cell types between the untreated and treated mice is calculated according to the metrics in Fig. \ref{ex vivo imaging}(c). Ground truth cell densities are determined using the microscope images by counting the fraction of cell nuclei (DAPI) labeled with the targeted probe (red and green channel). As the sensor does not have single-cell resolution, the cell density in the sensor images is determined by the fluorescence intensity in the tumor normalized by the area bounded by the dashed white lines in Fig. \ref{ex vivo imaging}(a) and (b). The background signal is mostly canceled out by measuring percent change.  

The quantified results from the sensor and microscope are shown in Fig. \ref{ex vivo imaging}(d). The sensor captures the general trends observed with the microscope, corresponding with the results in \cite{woo_cho_t_2022}. The increase in the density of CD8+ T-cells in both B16F10 samples (sensor: 847\%, microscope: 582\%) and the LLC samples (sensor: 38\%, microscope: 191\%) suggests an effector response to immunotherapy in both models. However, a larger increase in CD11b density after treatment in the LLC tumors (sensor: 66\%, microscope: 75\%) over the B16F10 tumors (sensor: 42\%, microscope: 51\%), suggests resistance in the LLC model due an increase in neutrophils. These trends would better reflect the results in \cite{woo_cho_t_2022} with a larger sample size to account for heterogeneity across the mice and neutrophil-specific biomarkers (CD11b also stains other myeloid cells). 

However, these results highlight the utility of multicolor fluorescence imaging in evaluating the response to cancer immunotherapy, enabling a differential measurement of both effector (e.g. CD8+ T-cell) and suppressor (e.g. neutrophil) populations. As shown by the increase in CD8+ T-cells in resistant LLC tumors, an increase in effector populations does not always correlate with response as the effector cells may be inhibited by suppressor cells. Therefore, simultaneously imaging suppressor populations such as neutrophils has two advantages: (1) enabling a more accurate assessment of response and (2) revealing the mechanisms of resistance (e.g. neutrophil interference with CD8+ T-cells) that can be targeted with second-line therapies (e.g. blocking T-cell-induced immunosuppressive inflammation signaling as done in \cite{woo_cho_t_2022}). Future \textit{in vivo} studies can highlight the unique capability of our sensor to analyze real-time dynamics in the spatial interactions of these populations, which is critical for developing a more nuanced understanding of the immune response \cite{vitale_intratumoral_2021}.
\begin{table*}[!t]
\captionsetup{justification=centering} 
\caption{COMPARISON OF STATE-OF-THE-ART CHIP-SCALE FLUORESCENCE IMAGE SENSORS}
\label{tab:comparison table}
\centering
\includegraphics[width=7.14in]{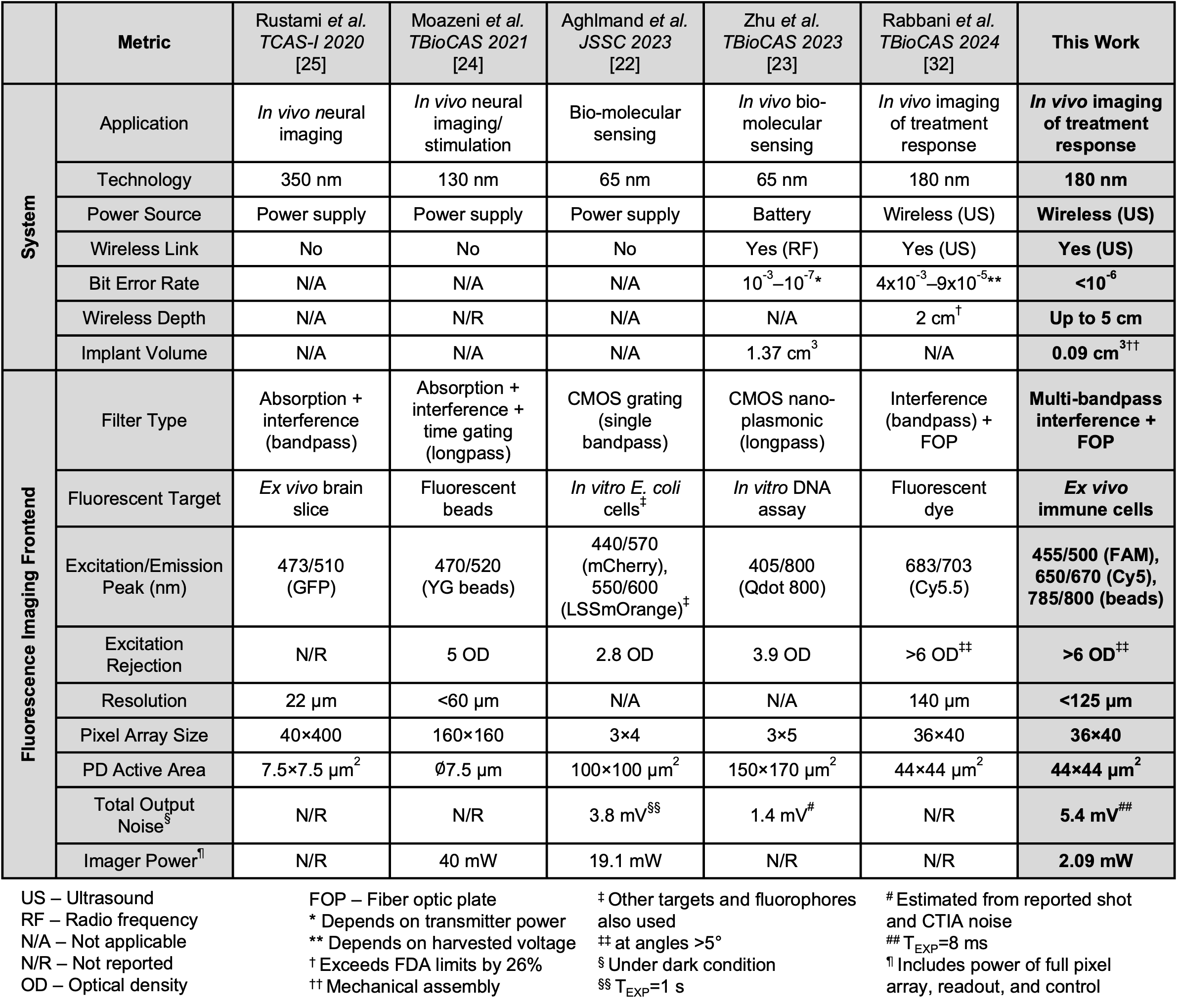}
\end{table*}
\section{Conclusion}
We present a fully wireless implantable image sensor capable of multicolor fluorescence imaging for real-time monitoring of response to cancer immunotherapy. A comparison of our work with recent chip-scale fluorescence imagers and sensors is shown in Table \ref{tab:comparison table}. To the knowledge of the authors, our work is the first to demonstrate fully wireless operation of the entire system with biologically relevant samples. In \cite{zhu_ingestible_2023}, a battery is used for power. In \cite{rabbani_towards_2024} the US link operates above FDA limits and low imager sensitivity limits wireless imaging to high concentrations of fluorescent dye. With a power harvesting frontend incorporating a cross-coupled charge-pump, we demonstrate safe operation at 5 cm depth in oil with US power densities at 31\% of FDA limits. The robust communication link demonstrates a BER better than 10\textsuperscript{-6} with a 13 kbps data rate. Moreover, optimization of the storage capacitor sizing enables a small form factor of 0.09 cm\textsuperscript{3} demonstrated with a mechanical assembly of the implant. 

Our system is specifically designed for multicolor fluorescence imaging with a three-channel laser driver to drive different color {\textmu}LDs, an US downlink for programming imaging and laser settings, and an optical frontend design consisting of a multi-bandpass interference filter and a FOP. Our optical frontend provides greater than 6 OD of excitation rejection of lasers within 15 nm of the filter band edge, a significant improvement over the CMOS metal filters reported in \cite{aghlmand_65-nm_2023,zhu_ingestible_2023} and competitive performance with the combination of absorption and interference filters in \cite{moazeni_mechanically_2021, rustami_needle-type_2020}. To the best of our knowledge, this work is the first chip-scale fluorescence imager capable of three-color imaging, which we demonstrate through imaging fluorescent beads. The pixel noise is on the same order of magnitude as \cite{aghlmand_65-nm_2023, zhu_ingestible_2023} despite these works using pixel sizes accommodating large low-noise readout circuits with higher power consumption. 

By imaging CD8+ T-cells and neutrophils populations in \textit{ex vivo} mouse tumors with or without immunotherapy, we show how multicolor fluorescence imaging can enable accurate identification of non-responders and their underlying resistance mechanisms. Such sub-millimeter imaging of multiple biomarkers is inaccessible to clinical imagers such as MRI, CT or PET and can inform personalized treatment regimens addressing the wide variability in response to immunotherapy across patients. With future work in biocompatible packaging and integration of optics for epi-illumination, our platform can open the door to real-time, chronic monitoring of the spatial interactions of multiple cell populations deep in the body.

\section*{Acknowledgments}
The authors would like to thank sponsors of BSAC (Berkeley Sensors and Actuators Center) and TSMC for chip fabrication. We appreciate technical discussion and advice from Prof. Rikky Muller, Efthymios Papageorgiou, Hossein Najafiaghdamand, and Mohammad Meraj Ghanbari. Thank you to Eric Yang, Jade Pinkenburg, and Kingshuk Daschowdhury for their technical assistance. Finally, we acknowledge Dr. Mohammad Naser from Biological Imaging Development CoLab (BIDC) and Kristine Wong from Laboratory for Cell Analysis (LCA) for the development of immunohistochemistry workflow and imaging. 

\printbibliography

 
\vspace{11pt}

\begin{IEEEbiography}[{\includegraphics[width=1in,height=1.25in,clip,keepaspectratio]{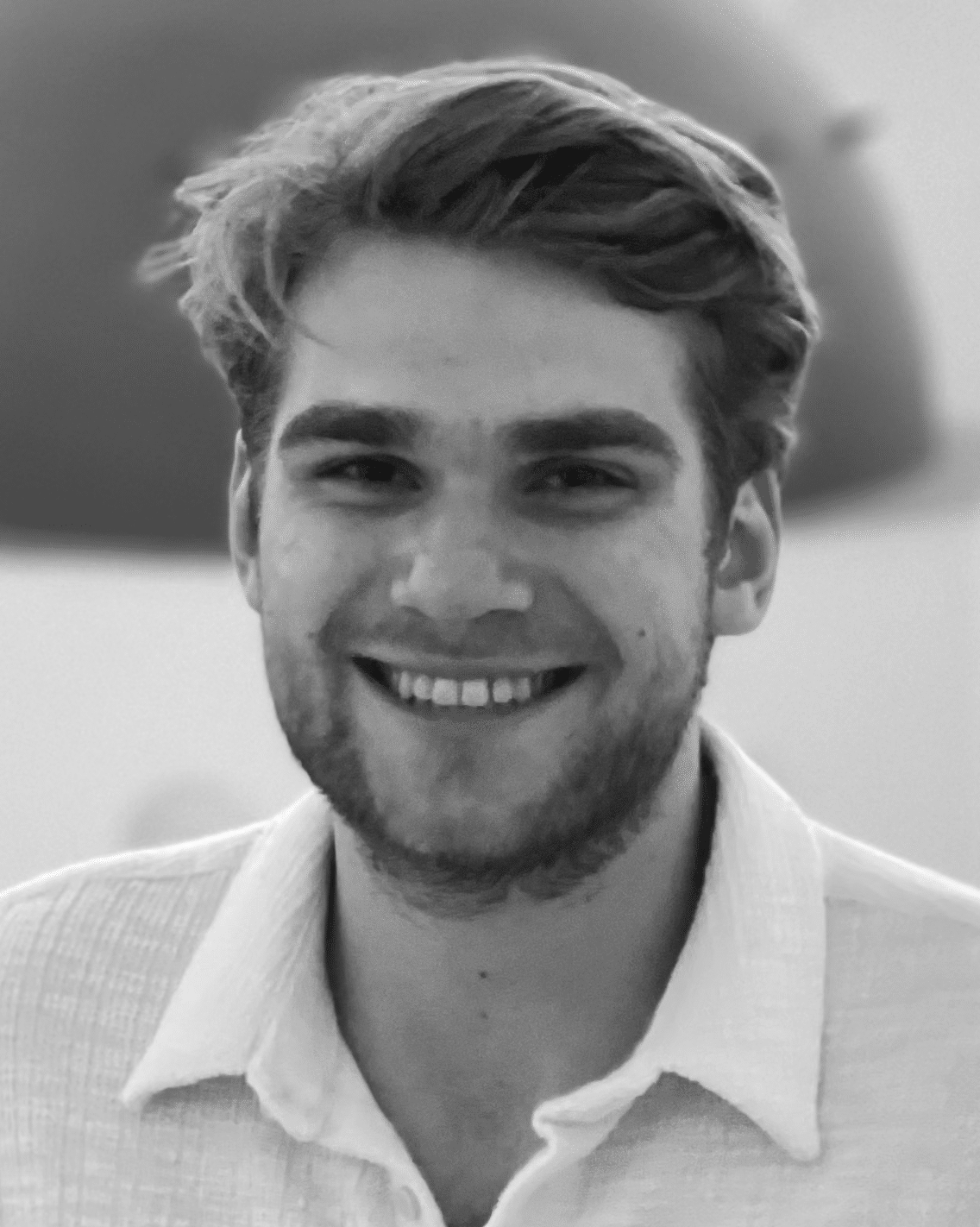}}]{Micah Roschelle}
(Graduate Student Member, IEEE) received his B.S. degree in electrical engineering from Columbia University, New York, NY, USA, in 2020. He is currently pursuing a Ph.D. in electrical engineering and computer sciences at the University of California, Berkeley, Berkeley, CA, USA. His research interests include implantable medical devices, lensless fluorescence imaging, and biomedical sensor design. 
\end{IEEEbiography}

\begin{IEEEbiography}[{\includegraphics[width=1in,height=1.25in,clip,keepaspectratio]{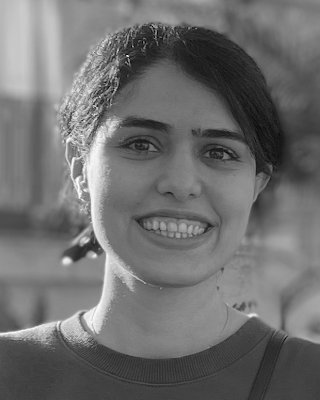}}]{Rozhan Rabbani}

(Graduate Student Member, IEEE) received the B.Sc. degree from Sharif University of Technology, Tehran, Iran, in 2018. She received her Ph.D. degree from the Department of Electrical and Computer Sciences, University of California Berkeley, Berkeley, CA, USA in 2024. At Sharif University of Technology, she worked on analog and mixed-signal circuit design to optimize power consumption for a wearable ECG sensor. She worked at Apple Inc. during Summers 2020 and 2022 working on calibration and test automation for high-speed applications. Her research at UC Berkeley was focused on developing biomedical circuits and sensors, specifically implantable image sensors for cancer therapy. She was the recipient of the Apple Ph.D. Fellowship in Integrated Circuits in 2022, the 2024 SSCS Rising Stars, and the 2024 SSCS Predoctoral Achievement Award.
\end{IEEEbiography}

\begin{IEEEbiography}[{\includegraphics[width=1in,height=1.25in,clip,keepaspectratio]{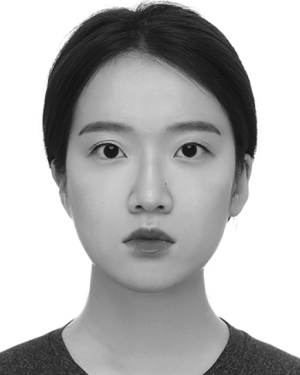}}]{Surin Gweon}
(Graduate Student Member, IEEE) received the B.S. degree in electrical engineering from Korea University, Seoul, South Korea, in 2018, and the M.S. degree in electrical engineering from the Korea Advanced Institute of Science and Technology (KAIST), Daejeon, South Korea, in 2020. She worked with System LSI Business, Samsung Electronics Company Ltd., Hwaseong, South Korea until 2023. She is currently pursuing the Ph.D. degree electrical engineering and computer sciences at the University of California at Berkeley (UC Berkeley), Berkeley, CA, USA. Her research interests include image sensor front-end and mixed-mode computing for implantable biomedical applications.
\end{IEEEbiography}

\begin{IEEEbiography}[{\includegraphics[width=1in,height=1.25in,clip,keepaspectratio]{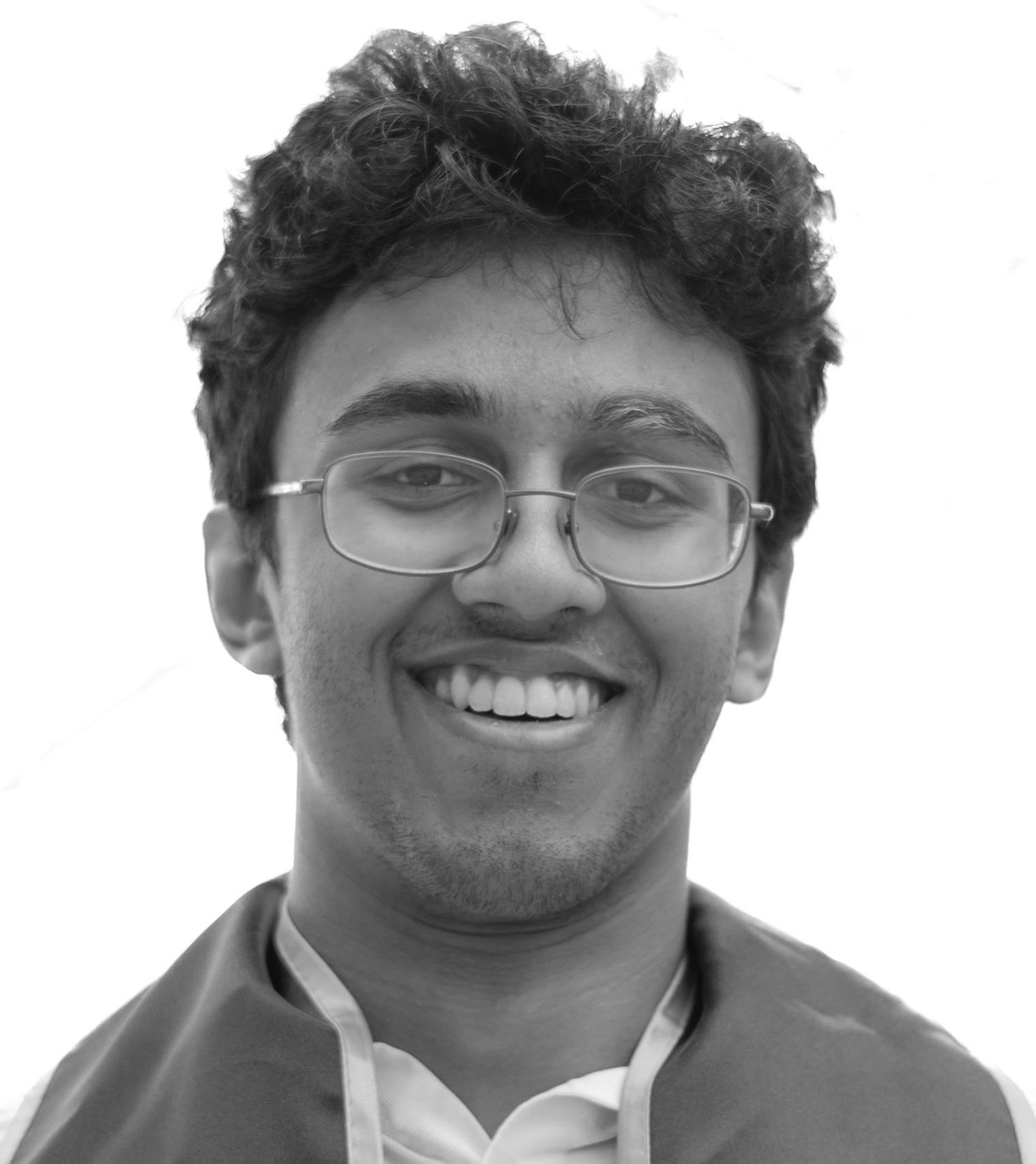}}]{Rohan Kumar}
(Graduate Student Member, IEEE) received his B.S. degree in electrical engineering and computer science (EECS) from the University of California, Berkeley, Berkeley, CA, USA, in 2024. He is currently pursuing a Ph.D. in EECS at UC Berkeley. His interests include electronic design automation, die-to-die interconnects, and open-source hardware.
\end{IEEEbiography}

\begin{IEEEbiography}[{\includegraphics[width=1in,height=1.25in,clip,keepaspectratio]{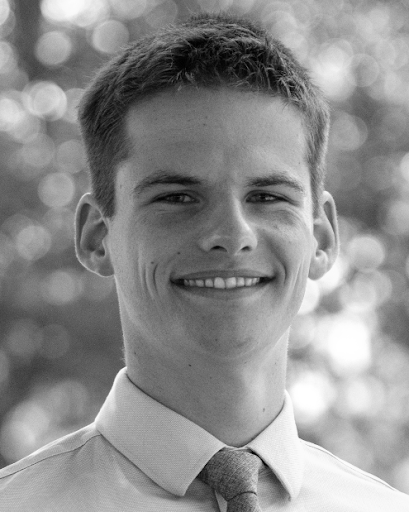}}]{Alec Vercruysse}
(Graduate Student Member, IEEE) received a B.S. degree in engineering from Harvey Mudd College in Claremont, CA, USA in 2023. He is currently pursuing a Ph.D. in electrical engineering and computer sciences at the University of California, Berkeley. His interests include the system-level design of circuits for implantable medical devices.
\end{IEEEbiography}

\begin{IEEEbiography}[{\includegraphics[width=1in,height=1.25in,clip,keepaspectratio]{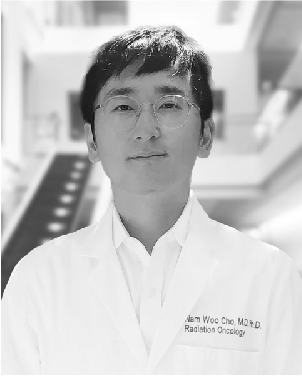}}]{Nam Woo Cho}
Dr. Nam Woo Cho, MD, PhD is a physician scientist in radiation oncology. Dr. Cho received his undergraduate degree from Harvard College, and MD/PhD degrees from the University of Pennsylvania. He completed internship in Internal Medicine at St. Mary’s Medical Center in San Francisco, and his residency in radiation oncology at UCSF. Following his postdoctoral work with Dr. Matthew Spitzer, he started his own research laboratory as an Assistant Professor in the Department of Radiation Oncology and Department of Otolaryngology-Head and Neck Surgery. His research focuses on understanding fundamental immunologic mechanisms that govern responses to immune stimulating therapies including radiation therapy and immune checkpoint inhibitors. Dr. Cho leverages molecular, cellular, organismal, and computational platforms to define novel mechanisms, pioneering the next generation of radio- and immune-therapeutics. 
\end{IEEEbiography}

\begin{IEEEbiography}[{\includegraphics[width=1in,height=1.25in,clip,keepaspectratio]{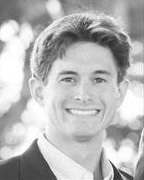}}]{Matthew H. Spitzer}
received the B.S. degree from Georgetown University, Washington, DC, USA and the Ph.D. degree from Stanford University, Stanford, CA, USA, in 2015. In 2016, he joined University of California San Francisco (UCSF), San Francisco, CA, USA as a UCSF Parker Fellow and a Sandler Faculty Fellow. He is currently Associate Professor in the Departments of Otolaryngology-Head and Neck Surgery and Microbiology \& Immunology at UCSF and an investigator of the Parker Institute for Cancer Immunotherapy, San Francisco, USA. His research aims to develop understanding of how the immune system coordinates its responses across the organism with an emphasis on tumor immunology by combining methods in experimental immunology and cancer biology with computation.
\end{IEEEbiography}

\begin{IEEEbiography}[{\includegraphics[width=1in,height=1.25in,clip,keepaspectratio]{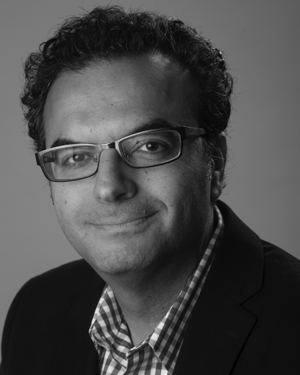}}]{Ali M. Niknejad}
(Fellow, IEEE) received the B.S.E.E. degree from the University of California at Los Angeles, Los Angeles, CA, USA, in 1994, and the master’s and Ph.D. degrees in electrical engineering from the University of California at Berkeley (UC Berkeley), Berkeley, CA, in 1997 and 2000, respectively. He is currently a Professor with the EECS Department, UC Berkeley, the Faculty Director of the Berkeley Wireless Research Center (BWRC), Berkeley, and the Associate Director of the Center for Ubiquitous Connectivity. His research interests include wireless and broadband communications and biomedical imaging and sensors, integrated circuit technology (analog, RF, mixed signal, and mm-wave), device physics and compact modeling, and applied electromagnetics. Prof. Niknejad and his coauthors received the 2017 IEEE Transactions on Circuits and Systems—I: Regular Papers Darlington Best Paper Award, the 2017 Most Frequently Cited Paper Award (2010–2016) at the Symposium on VLSI Circuits, and the CICC 2015 Best Invited Paper Award. He was a recipient of the 2012 ASEE Frederick Emmons Terman Award for his textbook on electromagnetics and RF integrated circuits. He was a co-recipient of the 2013 Jack Kilby Award for Outstanding Student Paper for his work on an efficient Quadrature Digital Spatial Modulator at 60 GHz, the 2010 Jack Kilby Award for Outstanding Student Paper for his work on a 90-GHz pulser with 30 GHz of bandwidth for medical imaging, and the Outstanding Technology Directions Paper at ISSCC 2004 for co-developing a modeling approach for devices up to 65 GHz. 
\end{IEEEbiography}

\begin{IEEEbiography}[{\includegraphics[width=1in,height=1.25in,clip,keepaspectratio]{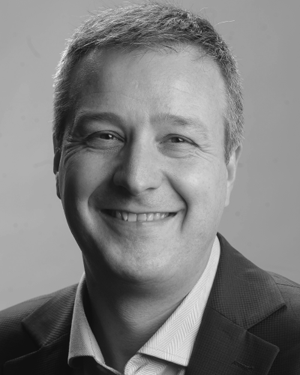}}]{Vladimir M. Stojanovi\'{c}}
(Fellow, IEEE) received the Dipl. Ing. degree from the University of Belgrade, Belgrade, Serbia, in 1998, and the Ph.D. degree in electrical engineering from Stanford University, Stanford, CA, USA, in 2005. He was with Rambus, Inc., Los Altos, CA, USA, from 2001 to 2004; and the Massachusetts Institute of Technology, Cambridge, MA, USA, as an Associate Professor, from 2005 to 2013. He is currently a Professor of electrical engineering and computer sciences with the University of California at Berkeley, Berkeley, CA, USA, where he is also a Faculty CoDirector of the Berkeley Wireless Research Center (BWRC). His current research interests include the design, modeling, and optimization of integrated systems, from CMOS-based VLSI blocks and interfaces to system design with emerging devices, such as NEM relays and silicon photonics, design and implementation of energy-efficient electrical and optical networks, and digital communication techniques in high-speed interfaces and high-speed mixed-signal integrated circuit (IC) design. Dr. Stojanović was a recipient of the 2006 IBM Faculty Partnership Award, the 2009 NSF CAREER Award, the 2008 ICCAD William J. McCalla, the 2008 IEEE TRANSACTIONS ON ADVANCED PACKAGING, and the 2010 ISSCC Jack Raper Best Paper and 2020 ISSCC Best Forum Presenter Awards. He was a Distinguished Lecturer of IEEE Solid-State Circuits Society from 2012 to 2013.
\end{IEEEbiography}

\begin{IEEEbiography}[{\includegraphics[width=1in,height=1.25in,clip,keepaspectratio]{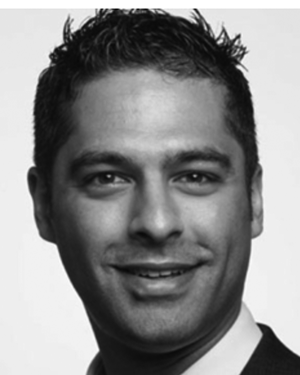}}]{Mekhail Anwar}
(Member, IEEE) received the B.A. degree in physics from the University of California Berkeley (UC Berkeley), Berkeley, CA, USA, where he graduated as the University Medalist, the Ph.D. degree in electrical engineering and computer sciences from the Massachusetts Institute of Technology, Cambridge, MA, USA, in 2007, and the M.D. degree from the University of California San Francisco (UCSF), San Francisco, CA, in 2009. In 2014, he completed a Radiation Oncology residency with UCSF. In 2014, he joined the faculty with the Department of Radiation Oncology, UCSF, with a joint appointment in Electrical Engineering and Computer Sciences at UC Berkeley (in 2021), where he is currently an Associate Professor. His research focuses on developing sensors to guide cancer care using integrated-circuit based platforms. His research centers on directing precision cancer therapy using integrated circuit-based platforms to guide therapy. His work in chip scale imaging has been recognized with awards from the DOD (Physician Research Award)  and the NIH (Trailblazer), and in 2020 he was awarded the prestigious DP2 New Innovator Award for work on implantable imagers. At UCB and UCSF he focuses on the development of implantable sensors across both imaging, molecular sensing and radiation therapy.  He is board certified in Radiation Oncology and maintains a clinical practice specializing in the treatment of GI malignancies with precision radiotherapy.
\end{IEEEbiography}

\vspace{11pt}

\vfill

\newpage
\appendix
\newpage

\begin{figure*}[thbp]
\centering
\includegraphics[width=7.13in]{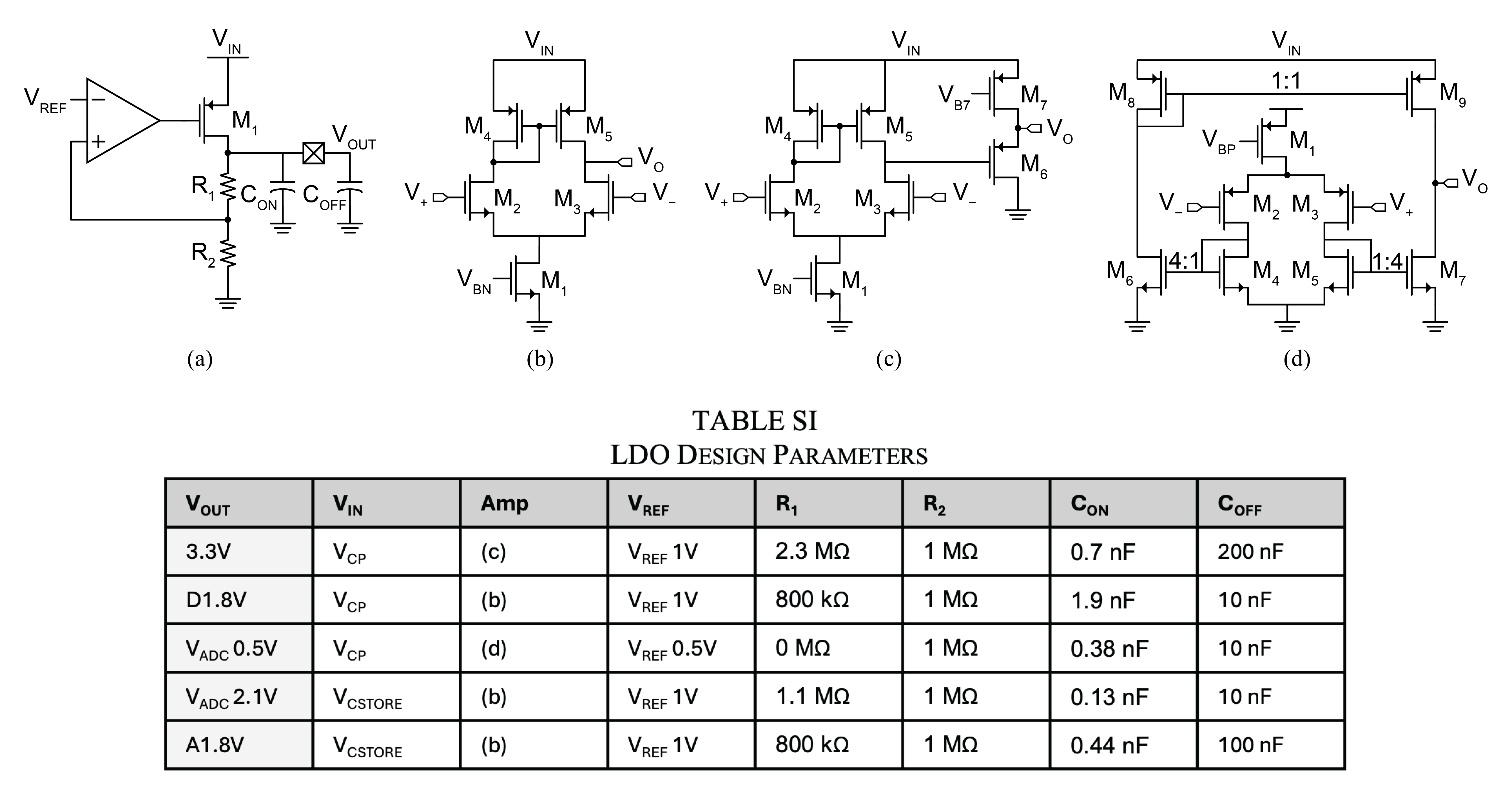}
\caption{(a) LDO schematic. (b–d) different error amplifier topologies. Table SI details design parameters for each of the 5 different LDOs.}
\label{LDOs}
\end{figure*}

\begin{figure}[thbp]
\centering
\includegraphics[width=3.49in]{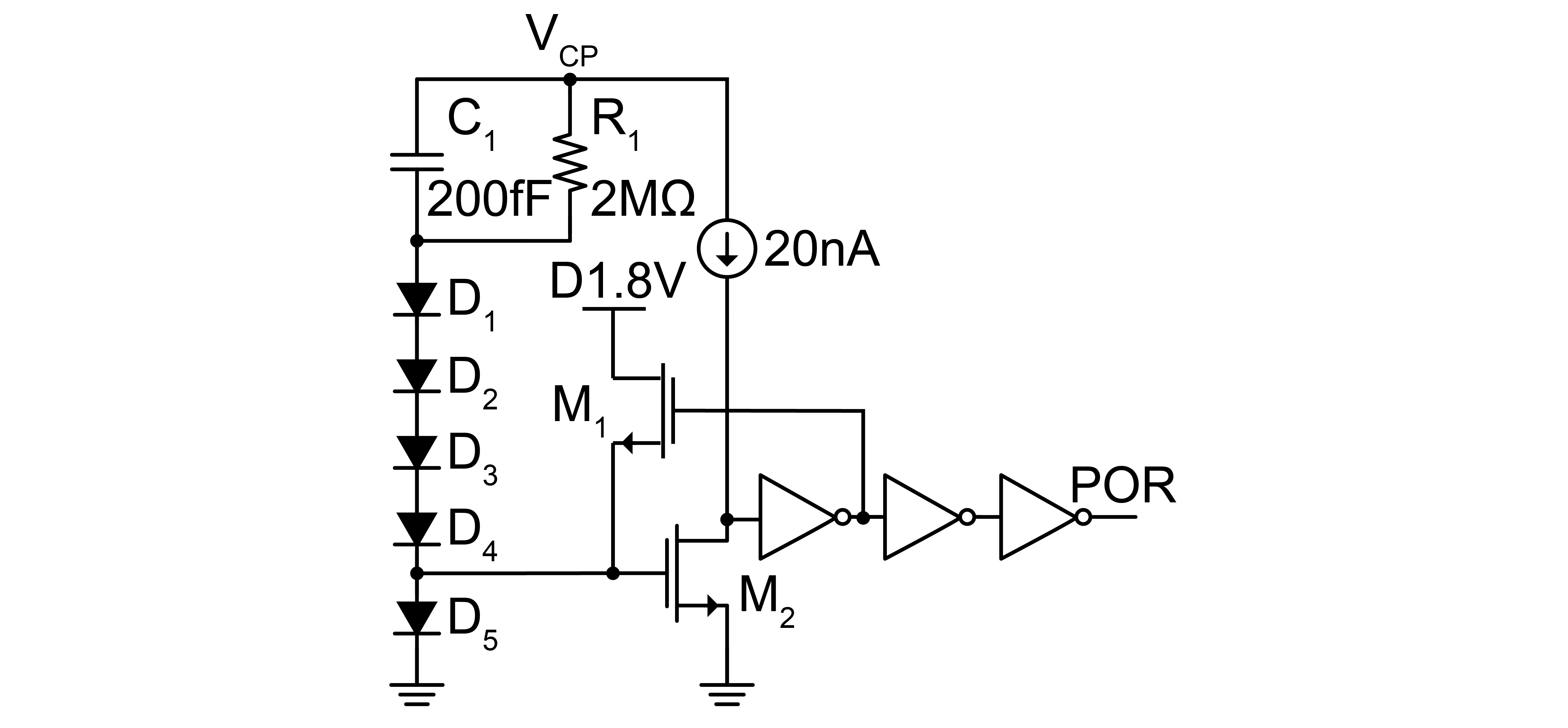}
\caption{Power-on reset (POR) circuit schematic. Initially, when the chip is charging up and V\textsubscript{CP} is below 3.9 V, the diodes, D\textsubscript{1}–D\textsubscript{5}, do not conduct current and the gate of M\textsubscript{2} is pulled low. Therefore, M\textsubscript{2} is off and the POR signal is pulled low, continuously resetting the finite state machine (FSM). When V\textsubscript{CP} reaches 3.9V, D\textsubscript{1}–D\textsubscript{5} turn on and the POR signal will be pulled high by M\textsubscript{2}, allowing the FSM to function normally. R\textsubscript{1} and C\textsubscript{1} prevent sudden fluctuations on V\textsubscript{CP} from triggering changes in the POR signal. As V\textsubscript{CP} falls below 3.9 V during the \textit{Readout} state, the feedback transistor, M\textsubscript{1}, ensures that M\textsubscript{2} stays on even as D\textsubscript{1}–D\textsubscript{4} turn off, maintaining a high POR signal as long as the digital 1.8 V LDO is still operational. The 3.9 V POR voltage is selected to be slightly higher than the 3.5 V minimum operational voltage on V\textsubscript{CP}, to ensure stable operation of the chip as the FSM wakes up.}
\label{POR}
\end{figure}

\begin{figure}[thbp]
\centering
\includegraphics[width=3.49in]{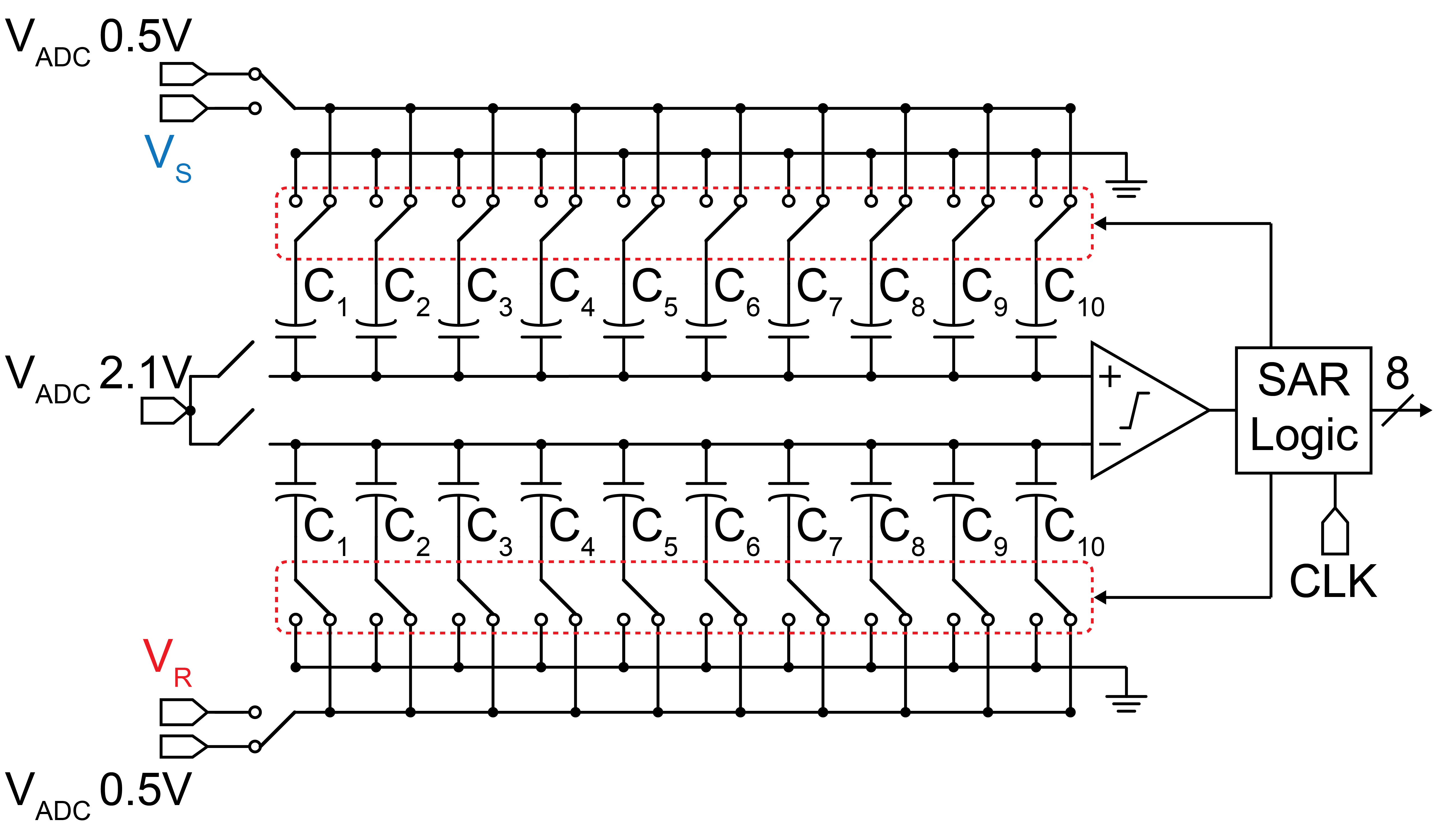}
\caption{ADC with conventional differential charge-redistribution SAR architecture. The ADC compares the pixel output voltage (V\textsubscript{S}) with the CDS reset voltage (V\textsubscript{R}) and produces an 8-bit digital pixel value. In the pixel architecture shown in Fig. 14(a), the output of the in-pixel CTIA is initially set to 0.6 V during the reset phase (T\textsubscript{RST}). The CTIA is supplied by the 1.8 V analog LDO such that its maximum output voltage is 1.6 V, resulting in a dynamic range of 1 V. During \textit{Readout}, the CDS outputs—pixel signal voltage (V\textsubscript{S}) and reset voltage (V\textsubscript{R})—are level-shifted up 1 V by the in-pixel source followers before the ADC samples them. Therefore, the common-mode input of the comparator is set to 2.1 V, which is provided by the V\textsubscript{ADC}2.1V LDO. This voltage is selected to be halfway between the minimum signal at the ADC input, corresponding to the CDS reset voltage (V\textsubscript{R}=1.6 V), and the maximum achievable pixel signal considering the CTIA dynamic range (V\textsubscript{S}=1.6+1 V=2.6 V).\newline \newline
Despite the 1 V headroom for the pixel output signal and given the typically low photocurrent signals from fluorescence imaging, the ADC dynamic range is set to 0.5 V, which dominates over the dynamic range of the CTIA. The ADC dynamic range is set through the V\textsubscript{ADC}0.5V LDO and is selected to achieve an LSB of 1.95 V, which adds negligent quantization noise to pixel readout noise (see Fig. 21(c) in the main text). This dynamic range is sufficient for capturing signals in the maximum exposure setting of 248 ms, where dark current uses up 338 mV of the dynamic range. \newline \newline
The strong-arm comparator utilizes PMOS input devices and operates on the 3.3 V supply with digital logic operating in the 1.8 V domain. The capacitive DAC (C\textsubscript{1}–C\textsubscript{10}) is implemented with MIM capacitors where the smallest capacitor, C\textsubscript{1} (40.56 fF), consists of two minimum-size unit capacitors of 20.28 fF each. \newline \newline
While the ADC has a 9-bit output, the 9\textsuperscript{th} sign bit is discarded and is not stored in the memory or transmitted via US backscatter. This sign bit is unnecessary because V\textsubscript{S} is always greater than V\textsubscript{R} even when no light is incident on the pixel: given the average pixel dark current of 14.9 fA and the 11 fF in-pixel integration cap, even at the shortest exposure time, T\textsubscript{EXP}=8 ms,  V\textsubscript{S} is expected to be 10.8 mV ({\textapprox}5.5 ADC LSBs) greater than V\textsubscript{R} on average. One conversion cycle of the ADC lasts 15 CLK cycles which is 16.3 {\textmu}s assuming a 920 kHz CLK from the US carrier. The simulated effective number of bits (ENOB) of the extracted ADC with the buffers, LDOs and PTAT is 7.93 bits.}
\label{ADC}
\end{figure}

\begin{figure}[thbp]
\centering
\includegraphics[width=3.49in]{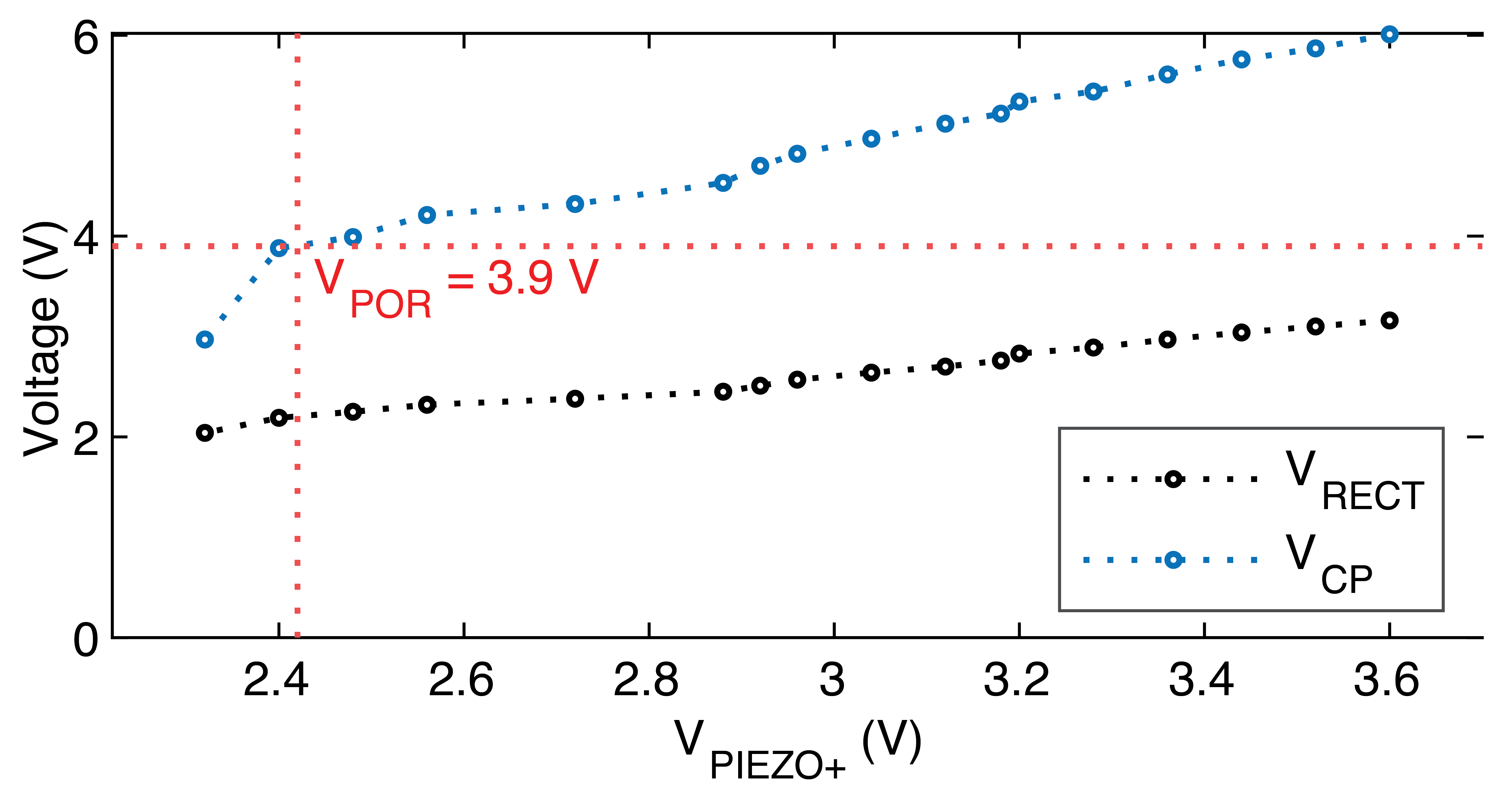}
\caption{Measured harvested voltage at the output of the rectifier (V\textsubscript{RECT}) and the output of the charge pump (V\textsubscript{CP}) for different voltages at the input of the rectifier (V\textsubscript{PIEZO+}). A minimum voltage of V\textsubscript{PIEZO+}=2.42 V is required to harvest V\textsubscript{CP}=3.9 V, high enough to trigger the POR (Fig. S2) and, thus, ensure stable operation of the ASIC. This voltage is 27\% lower than for nominal operation of the chip (V\textsubscript{CP}=5.5 V), which requires V\textsubscript{PIEZO+}=3.3 V. V\textsubscript{RECT} is less than V\textsubscript{PIEZO+} due to the nonzero |V\textsubscript{DS}| of the actively controlled PMOS switches (M\textsubscript{3} and M\textsubscript{4} in Fig. 9(a)) in the active rectifier.}
\label{rectifier char}
\end{figure}

\begin{figure}[thbp]
\centering
\includegraphics[width=3.49in]{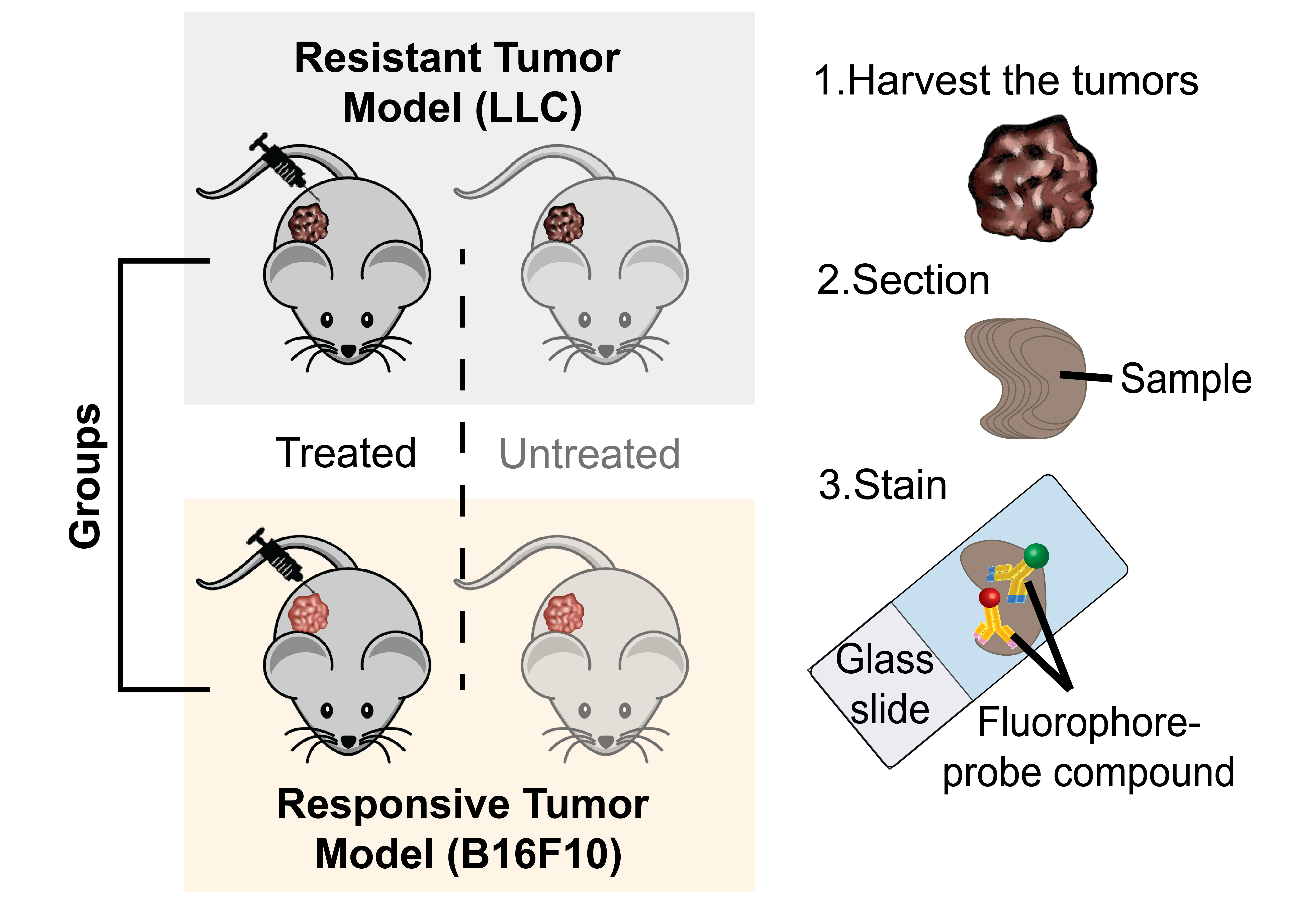}
\caption{Experimental design for the \textit{ex vivo} mouse experiment.}
\label{ex vivo setup}
\end{figure}

\end{document}